\documentclass{aa}

\usepackage{natbib}
\usepackage{lscape}
\usepackage{txfonts}
\usepackage{rotating}
\usepackage{longtable}

\begin{document}


\title{Chromospheric Activities and Kinematics for Solar Type Dwarfs and Subgiants: Analysis of the Activity Distribution and the AVR}


\author{J.S. Jenkins\inst{\ref{inst1},\ref{inst2}}\and F. Murgas\inst{\ref{inst3}}\and P. Rojo\inst{\ref{inst1}}\and H.R.A. Jones\inst{\ref{inst2}}\and 
A.C. Day-Jones\inst{\ref{inst1},\ref{inst2}}\and M.I. Jones\inst{\ref{inst1},\ref{inst4}}\and J.R.A. Clarke\inst{\ref{inst5}}\and M.T. Ruiz\inst{\ref{inst1}}\and D.J. Pinfield\inst{\ref{inst2}}}
\institute{Departamento de Astronom\'ia, Universidad de Chile, Camino del Observatorio 1515, Las Condes, Santiago, Chile\email{jjenkins@das.uchile.cl}\label{inst1}\and 
Center for Astrophysics Research, University of Hertfordshire, College Lane, Hatfield, Herts, AL10 9AB, UK\label{inst2}\and 
Instituto de Astrof\'isica de Canarias, Via Lactea, E38205, La Laguna, Tenerife, Spain\label{inst3}\and 
European Southern Observatory, Alonso de C{\'o}rdova 3107, Casilla 19001, Vitacura, Santiago 19, Chile\label{inst4}\and 
Departamento de F\'isica y Astronom\'ia, Facultad de Ciencias, Universidad de Valpara\'iso, Ave. Gran Breta\~na 1111, Playa Ancha, Casilla 5030, Valpara\'iso, Chile\label{inst5}}
\thanks{Based on observations made with the ESO telescopes at the La Silla Paranal observatory under programme ID's 076.C-0578(B), 077.C-0192(A), 082.C-0446(A) and 
082.C-0446(B).}

\date{Received 15 December 2010: Accepted --}


\begin{abstract}
{}
{In this work we present chromospheric activity indices, kinematics, radial-velocities and rotational velocities for more than 850 FGK-type dwarfs and 
subgiant stars in the southern hemisphere and test how best to calibrate and measure $S$-indices from echelle spectra.} 
{We measured our parameters using high resolution and high S/N FEROS echelle spectra acquired for this purpose.} 
{We confirm the bimodal distribution of chromospheric activities for such stars and highlight the role 
that the more active K-dwarfs play in biasing the number of active stars.  We show that the age-activity relationship does appear to continue to ages older than the Sun 
if we simply compare main sequence stars and subgiant stars, with an offset of around 2.5~Gyrs between the peaks of both distributions.  Also we show evidence 
for an increased spin-down timescale for cool K dwarfs compared with earlier F and G type stars.  

We highlight that activities drawn from low resolution spectra ($R<2'500$) 
significantly increase the rms scatter when calibrating onto common systems of measurements, like the Mt.~Wilson system.  Also we show that the older 
and widely used catalogue of activities in the south compiled by Henry et al. (1996) is offset by more recent works at the $\sim$0.1~dex level 
in log$R'_{\rm{HK}}$ through calibrator drift.  

In addition, we show how kinematics can be used to preselect inactive stars for future planet search projects.  We see the well known trend between 
projected rotational velocity and activity, however we also find a correlation between kinematic space velocity and chromospheric activity.  
It appears that after the Vaughan-Preston gap there is a quick step function in the kinematic space motion towards a significantly larger spread in velocities.  We speculate on reasons for this 
correlation and provide some model scenarios to describe the bimodal activity distribution through magnetic saturation, residual low level gas accretion or 
accretion by the star of planets or planetesimals.  
Finally, we provide a new empirical measurement for the disk heating law, using the latest age-activity 
relationships to reconstruct the age-velocity distribution for local disk stars.  We find a value of 0.337$\pm$0.045 for the exponent of this power law  (i.e. $\sigma_{\rm{tot}} \propto$~t$^{0.337}$),
 in excellent agreement with those found using isochrone fitting methods and with theoretical disk heating models.}
{}

\end{abstract}



\titlerunning{Chromospheric Activities and Kinematics}
\authorrunning{J.S. Jenkins et al.}

\maketitle



\section{Introduction}

The characterisation of bright stars in the solar vicinity is of primary importance for better understanding the formation and evolution 
of these objects.  Such fundamental characterisation is also necessary for a more complete understanding of stellar populations and galactic 
structure.  Studies of stellar chromospheric activity in large samples of nearby stars (\citealp{duncan91}; \citealp{henry96}; \citealp{wright04}; \citealp{jenkins06c, jenkins08}) \rm
have found that the magnetic activity of solar type dwarfs and subgiants (SG) follows a 
bimodal distribution.  In addition, the activity of these stars follows similar long term cycles like the Sun's 11~yr activity cycle and magnetic 
activity drops rapidly for the first $\sim$Gyr of a star's life and then tails off more slowly afterwards.  These previous studies provide good statistics on 
the nature of Sun-like magnetic activity.

Another interesting aspect of stellar population studies is the bulk kinematic space motion of stars in the solar neighbourhood.  The 
Hipparcos spacecraft has allowed precision proper motions and parallaxes of nearby stars to be used, along with spectroscopic observations that produce 
accurate kinematic space motions for large samples of stars.  Studies such as that from \citet{dehnen98} or \citet{murgas10} confirmed kinematical structures like 
moving groups exist in the galaxy, although other authors claim such kinematic over-densities are not young group stars but have a dynamical origin (e.g. 
\citealp{antoja08}).  Most recently, the large Geneva-Copenhagen Survey (GCS) of stars within 40~pc of Sun have provided a wealth of kinematic data and revealed 
the kinematic space component distributions to a significance much larger than any that came before (\citealp{nordstrom04}; \citealp{holmberg07}).

In both of the above areas of astrophysics it is necessary to maintain a clear homogeneous methodology and observational 
strategy.  Performing observations on large samples with the same instrumental setup, observing methods and data reductions limits any random errors and adds a significant level of 
robustness to the results.  We present the continuation of our project to better understand the properties of solar-type stars 
through analysis of high resolution spectra of a large sample of bright stars in the southern hemisphere.  In this work we outline the 
chromospheric activities and kinematic motions we have measured for our large database of high resolution and high S/N solar type stars in the southern hemisphere.  \rm

\section{Observations \& Reduction}

\begin{figure}
\vspace{4.5cm}
\hspace{-4.0cm}
\includegraphics{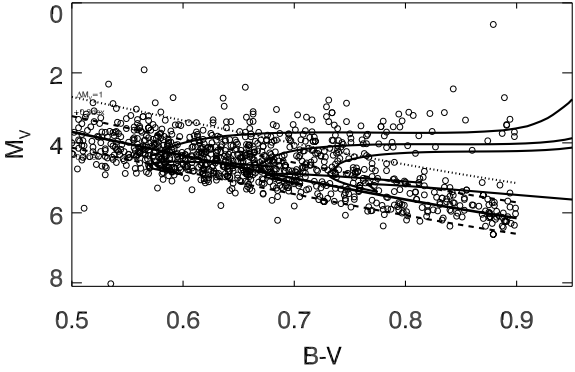}
\vspace{0.5cm}
\caption{A Hipparcos HR-diagram for our entire sample.  The three solid curves are Y2 isomass tracks set at a solar mass and with three different metallicities of -0.3, 0.0 and +0.3 dex 
running from left to right.  These tracks highlight the changing position of the main sequence as a function of metallicity and we highlight this by dashed lines on either side of the thick solid 
Hipparcos main sequence.  The dotted line is our cut to separate true subgiant stars from super metal-rich main sequence stars.}
\label{feros_hr}
\end{figure}

Our total sample is shown in Fig.~\ref{feros_hr} plotted on a colour-magnitude diagram by open circles.  The solid curves represent three different 
Y2 mass tracks (\citealp{demarque04}) showing the evolutionary change with metallicity (-0.3, 0.0, +0.3~dex) for a solar mass star.  Our sample in this plot 
ranges in age from below 1~Gyr to over 10~Gyrs, with the bulk of the sample distribution located between 5-10~Gyrs.  We note we have made 
first pass estimates for the masses of our sample.  The monotonically sloped and thick solid line, 
along with associated flanking dashed lines, represent the Hipparcos main sequence (MS) and the estimated +/-0.3~dex [Fe/H] scatter in the sequence.  The dotted line 
represents a cutoff between MS stars and SG stars.  This was determined using the distance from the Hipparcos MS as a function of 
$B-V$ colour where we set all SGs to have distance from the MS of 1 or greater.  The full selection details are explained in \citet{jenkins08}, 
hereafter referred too as Paper I.  Therefore, this plot merely serves to highlight the sample distribution we will discuss in the chapters that follow.

All target stars and calibration data were observed using the Fibre-fed Extended Range Optical Spectrograph (FEROS)
mounted on the MPG/ESO - 2.2m telescope on the La Silla site in Chile over four observing runs.  The first two runs were 
three nights in length, from 02 to 05 of February and September 2006 and the second two runs ran from 07 to 12 of October 2008 and 
from 13 to 17 of March 2009.  Over the 16 night observing program we compiled over 950 spectra with S/N ratios of 
100$-$200 in the continuum at 7500\AA, (necessary for the accurate radial-velocities (RVs) and for future atomic abundance work), and a median S/N $\sim$60 
at the CaHK lines (3955\AA) which is sufficient for accurate chromospheric activity analyses, and all at a resolving power of $R\sim$46$'$000.  
All calibration files needed for reducing the stellar spectra (flat-fields, bias and arc frames) were
obtained at the beginning and end of each night's observing following the standard ESO calibration plan.

The reduction procedure for all spectra is described in detail in Paper I and therefore 
the description is given here in short.  All data were debiased, flatfielded, had the scattered-light removed, optimally extracted and 
had the blaze function removed using the FEROS pipeline and a number of Starlink procedures.  Note that as in Paper I the data were binned to
a linear wavelength step of 0.03\AA /pixel, or a mean velocity resolution of 1.42kms$^{-1}$/pixel, and shifted into the rest frame by 
cross-correlating with an observation of HD102117 (G6V), which has been shown to exhibit a RV variation $<$20ms$^{\rm{-1}}$ 
(\citealp{jones02b}), and has been shifted into the rest frame.

\section{Chromospheric Activities}

The chromospheric activities were derived using the method explained in Paper I, however new calibration stars were used to convert these FEROS S-indices to the 
Mt.~Wilson system of measurements.  The new calibration stars were included to increase the accuracy of the calibration procedure given that the work 
of \citet{baliunas95} and \citet{lockwood07} have published a number of long term activity stable stars i.e. stars that show no significant long term activity 
zero point drift.  We tried to include as many of the southern stable stars from these works as possible.  The 
calibration stars and their derived values are shown in Table~\ref{tab:act_cals}.

In brief, the indices were measured by calculating the flux in the line cores of the Ca\sc~II\rm~H and K lines using 
a triangular bandpass, and comparing this flux to the flux in square bandpasses at either side of the calcium doublet.  The fit to the calibration stars used to convert to the 
Mt.~Wilson system (\citealp{duncan91}), hereafter referred to as MW, system of measurements is discussed in $\S$3.2, where we employ the 1.09\AA\ triangular core bandpasses for our 
operating resolving power of $\sim$46$'$000 since this synthetically mimics the MW methodology.

\begin{figure}
\vspace{4.5cm}
\hspace{-4.0cm}
\includegraphics{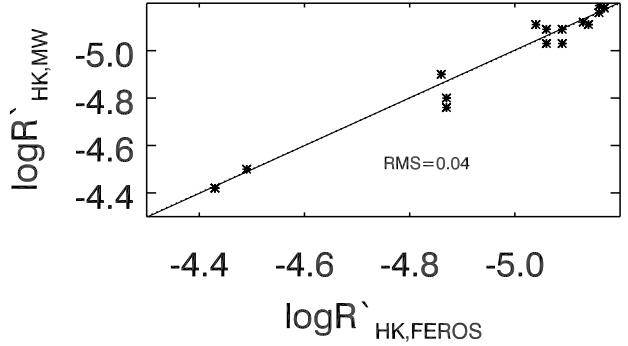}
\vspace{0.5cm}
\caption{Calibrated log$R'_{\rm{HK}}$ activity values onto the Mt. Wilson system of measurements.  The solid line marks the best straight line fit to the 
calibration data and the total rms scatter of 0.04 is shown.  Here the 1:1 and best fit are indistinguishable.  Given the intrinsic 
activity scatter for dwarf stars the calibration is proven to be robust.}
\label{rhk_fit}
\end{figure}

Figure~\ref{rhk_fit} shows the final log$R'_{\rm{HK,FEROS}}$ activity indices for the calibrators against their associated MW measurements (log$R'_{\rm{HK,MW}}$), where each 
star's chromospheric flux has been isolated by normalising to the 
bolometric luminosity of the star and removing the star's photospheric contribution.  The dotted line represents the 1:1 relationship with the solid line the best 
straight line fit to the data.  Clearly these log$R'_{\rm{HK}}$ indices have no residual slope or offset compared to the MW values and should provide highly accurate 
and robust activity measurements.  The rms we found around the best fit is 0.04~dex, which is slightly above the 0.034~dex rms found in Paper I.  This is mainly due to the low number of 
calibrators used in these works and the increased baseline of observations.  All stars, activity values, along with magnitudes and colours are listed in columns 1-5 of Table~4.

\subsection{Comparison with Other Work}

\begin{figure}
\vspace{4.5cm}
\hspace{-4.0cm}
\includegraphics{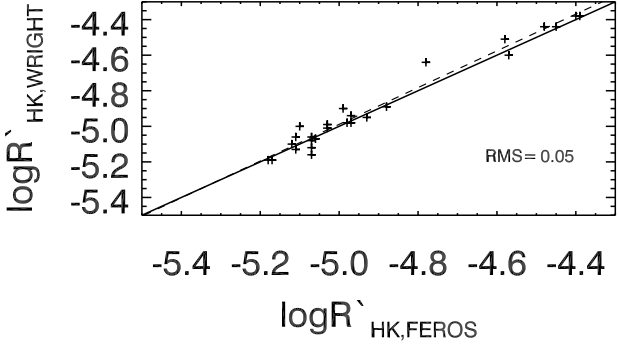}
\vspace{0.5cm}
\caption{Comparison between our FEROS chromospheric activities and those in Wright et al. (2004).  The rms scatter here is close to the intrinsic scatter found in the 
calibration method (0.05).  The dashed 1:1 and the best fit straight line overlap significantly and there is no differential trend found between the two.}
\label{rhk_fer_wright}
\end{figure}

Due to the low number of calibrators we used, particularly around the Vaughan-Preston (VP) gap region (\citealp{vaughan80}), comparisons with other works containing more overlapping stars can 
help to fully test if these values are accurate over a wider range.  Fig.~\ref{rhk_fer_wright} shows our measured chromospheric activities against values derived by \citet{wright04}.  As 
in Paper I, no offset or trend is found between the two works and the rms around the best fit is only 0.05~dex, which given the intrinsic variability of dwarf stars (can be at the level of 
$\pm$0.15~dex; \citealp{jenkins06c}), highlights the accuracy of both techniques.  Since we both use high resolution and high S/N datasets, along with employing a similar methodology to 
closely resemble the MW method, such a low rms might be expected.

\begin{figure}
\vspace{7.5cm}
\hspace{-4.0cm}
\includegraphics{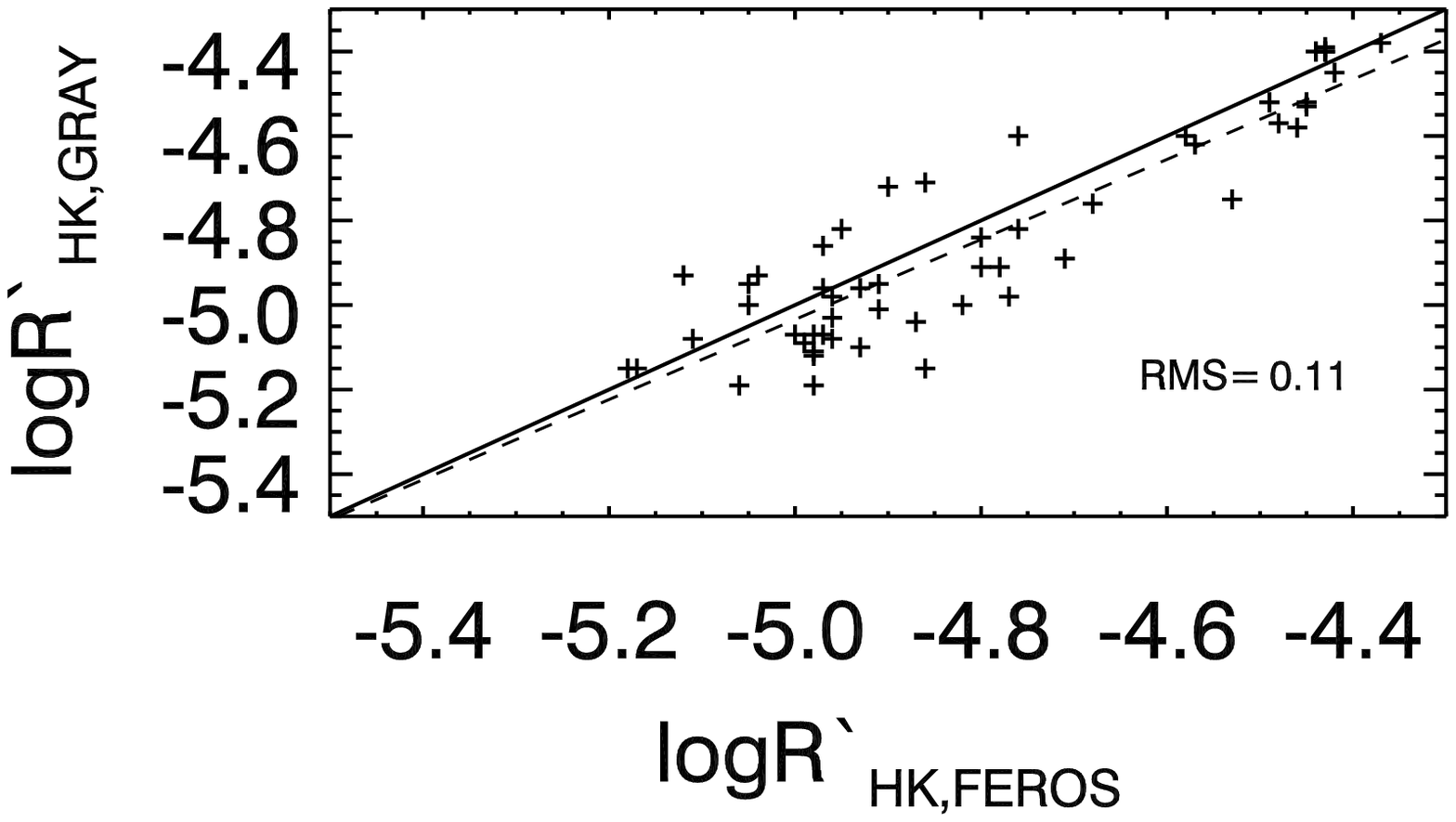}
\includegraphics{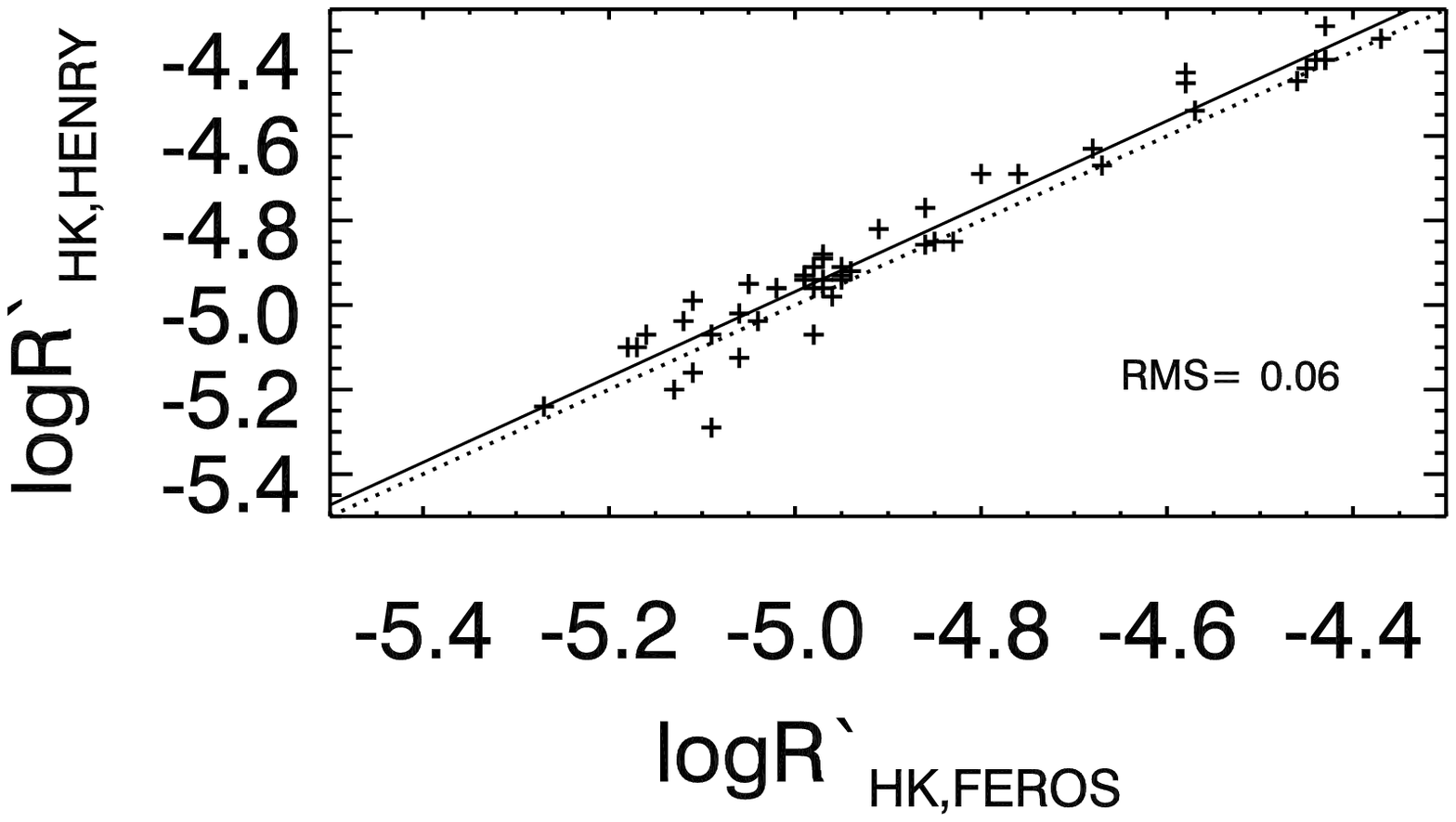}
\vspace{2.0cm}
\caption{Comparison between our FEROS chromospheric activities and those in Gray et al. (2006) is shown in the top panel.  There is a large amount of scatter in this plot (0.11), which is 
significantly 
larger than the intrinsic rms we found from our calibration stars (0.04).  However, there is no significant trend, as shown by the best fit straight line (solid line) compared to the dashed 1:1 
relationship.  The lower panel shows the same comparison but against the published values of \citet{henry96} where an offset is apparent.  The offset only seems apparent for moderate to 
active stars.}
\label{rhk_fer_gray}
\end{figure}

In Paper I we found that our log$R'_{\rm{HK}}$'s were offset from the indices derived by \citet{gray06} by a factor of -0.15~dex.  
When cross-matching the Gray et al. southern sample with that of Henry et al. we were able to better highlight this offset, with the Gray et al. values clearly systematically offset by -0.15~dex.  
Fig.~\ref{rhk_fer_gray} (top panel) shows our new measured log$R'_{\rm{HK}}$ indices against those from Gray et al. and now with more data we find a small and non-significant offset in the 
opposite 
direction (+0.05~dex).  The dashed line marks the 1:1 relationship and the solid line represents the best straight line fit to the data.  The best fit highlights the small offset between the 
two datasets but it is also interesting to note that there appears to be two populations, with most stars having Gray et al. log$R'_{\rm{HK}}$ indices below the 1:1 relationship and a small 
tail of stars significantly above the 1:1 relationship.  There is also a lack of stars between these two extremes.  This gap may arise simply because of the small number of overlapping stars between 
Gray et al. and this work.  The rms around the best fit (0.11~dex) is significantly larger than that from the Wright et al. data and also is much larger than the rms scatter around 
the MW calibration measurements.  This likely highlights the fact that 
the Gray et al. data were observed at significantly lower resolution ($R <$~2$'$000) than the data presented here, with this level of scatter could mask any zero-point offsets in the data.

The lower panel in Fig.~\ref{rhk_fer_gray} shows our newly calibrated data against the 1996 values from Henry et al. where there appears a clear offset once the log$R'_{\rm{HK}}$ 
values are larger than -4.95~dex.  To test the significance of this offset we apply a Sign-test to the data.  This reveals that the total mean of our dataset, compared with 
Henry et al., is significantly different at greater than the 3$\sigma$ level.  We believe this difference is purely from calibrator drift, since we did not see any offset in Paper I when we used less 
long-term stable calibrators.  In comparison the Sign-tests for the Gray et al. and Wright et al. datasets are significantly different.  The Gray et al. data also show weak evidence for a possible 
mean offset difference, but with much lower significance than Henry et al.  Also the Wright et al. data show no evidence for any offset at all.

We have also compared the values of Henry et al., Wright et al. and Gray et al. among themselves to confirm the offsets we see and also to highlight the reliability of each method.  The 
Sign-test between Henry et al. and Wright et al. confirmed the offset we find at greater than 5$\sigma$ and it is even more pronounced between Henry et al. and Gray et al., not surprising 
since we found Gray et al. may be offset against our data in the opposite direction to that of Henry et al.  Finally the Sign-test revealed there was essentially no offset between the activities 
in Wright et al. and those in Gray et al.  These tests reveal the importance of defining a set of long term stable chromospheric calibrator stars to better compare past and present data sets.

\subsection{Chromospheric Activity Calibration Tests}

\begin{figure*}
\vspace{5.5cm}
\hspace{-4.0cm}
\includegraphics{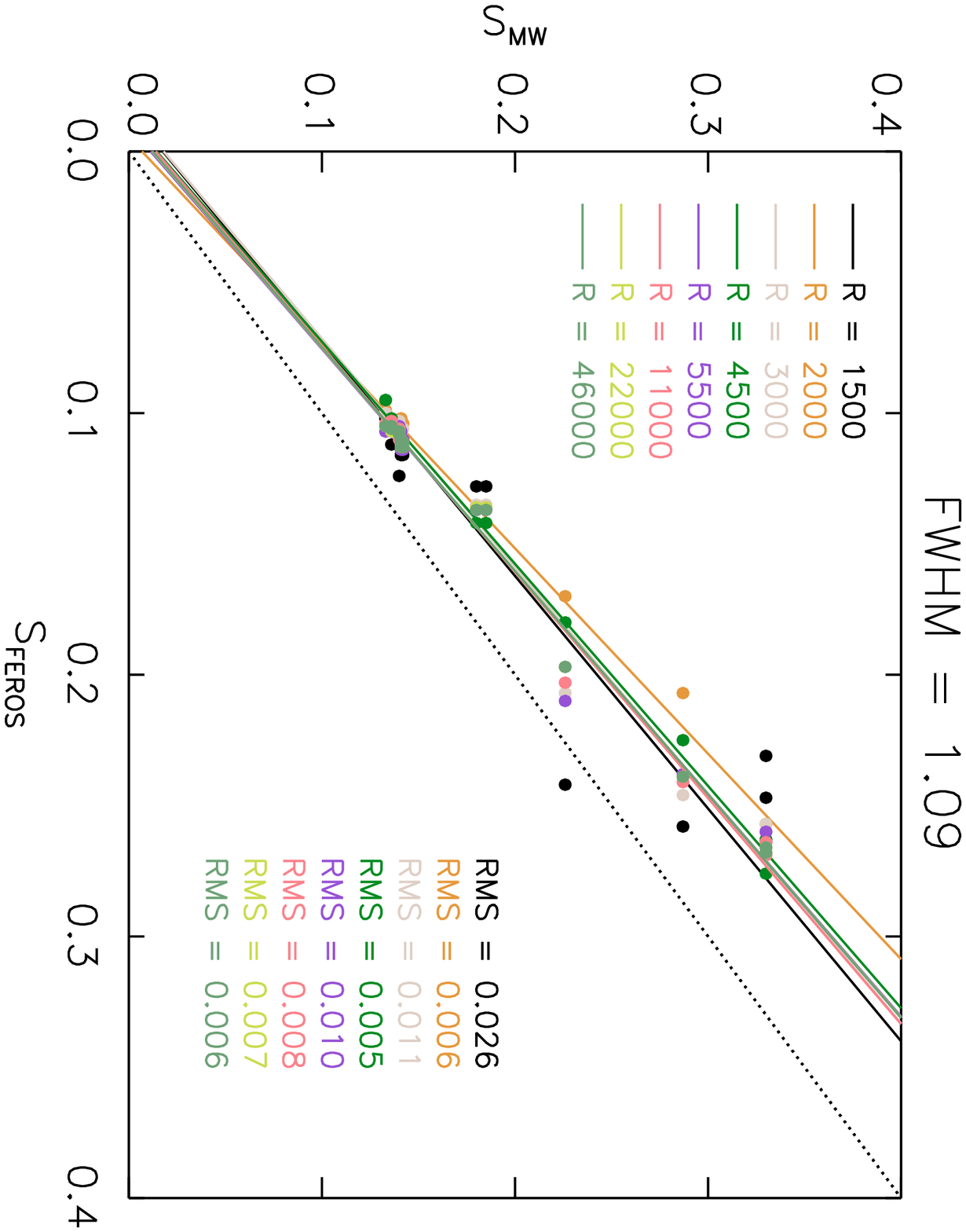}
\includegraphics{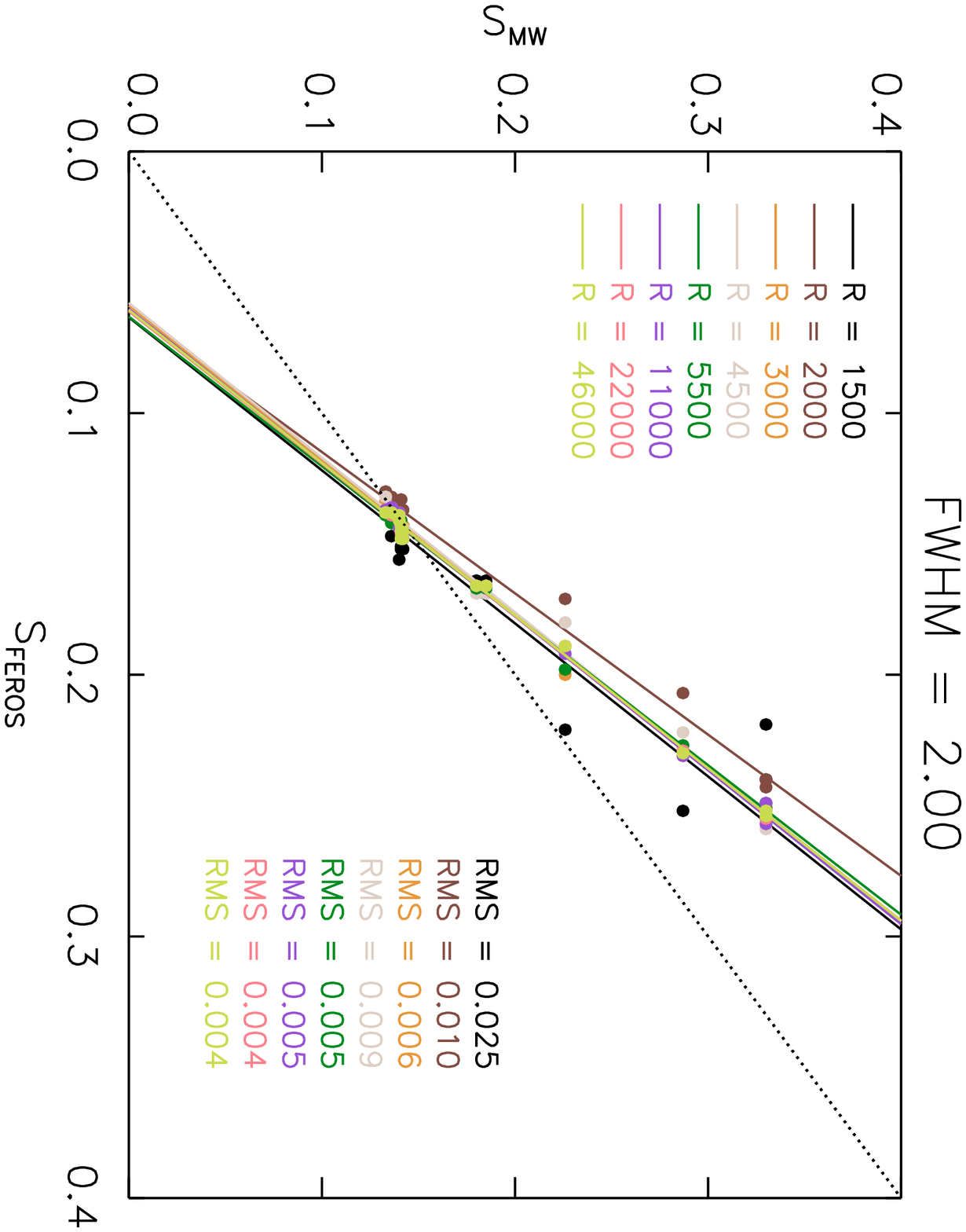}
\includegraphics{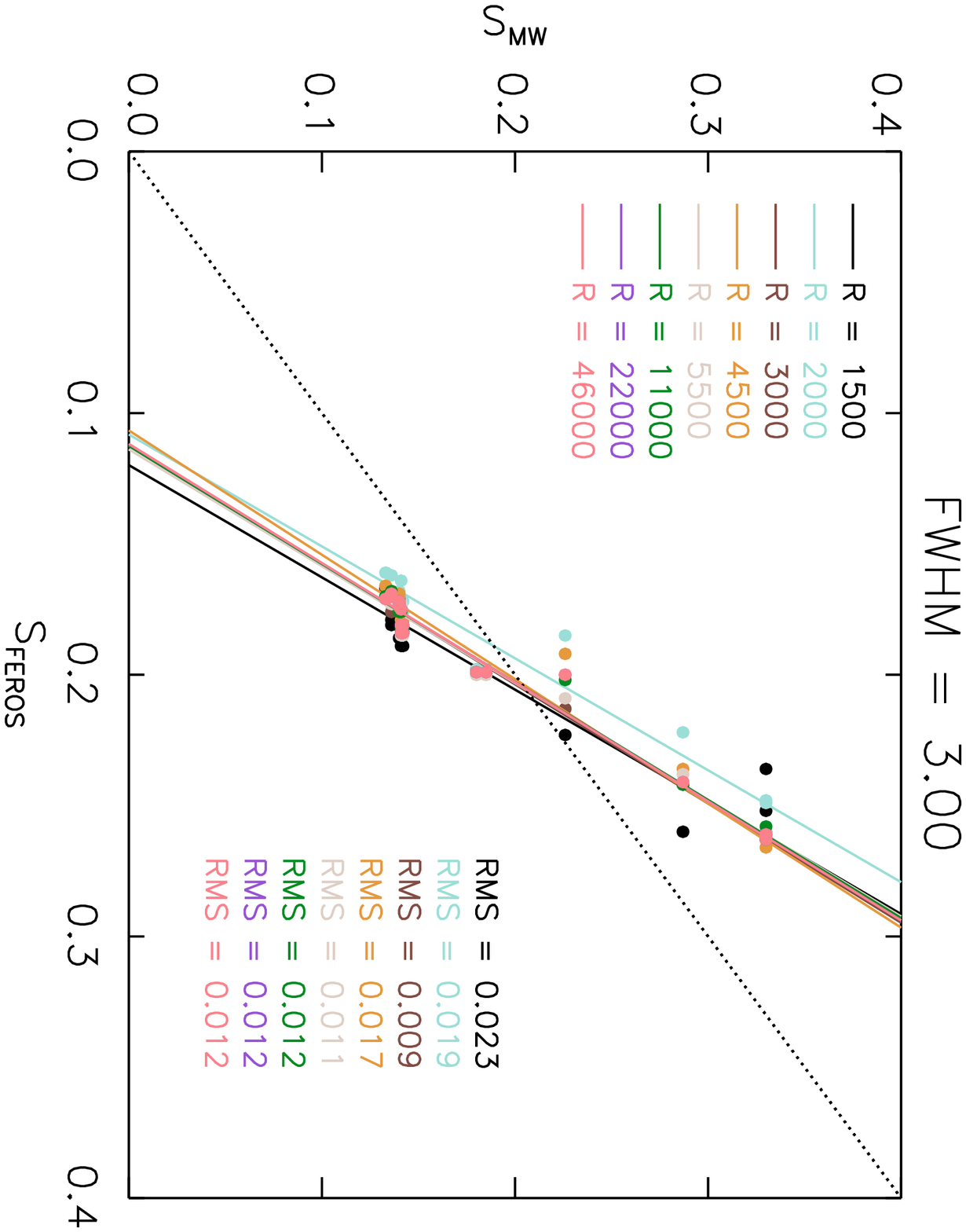}
\includegraphics{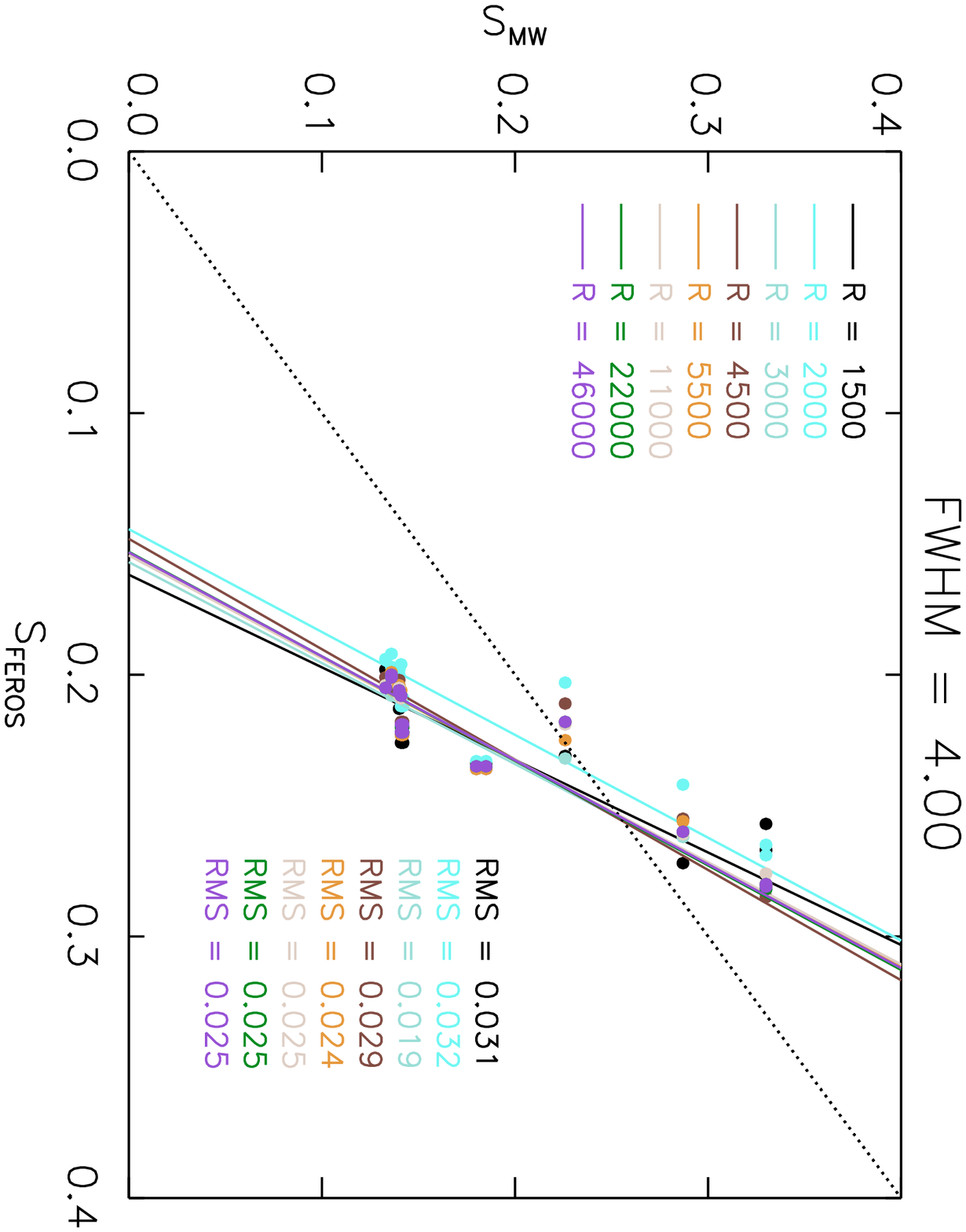}
\vspace{6.0cm}
\caption{Four separate calibration panels representing four different triangular bandpass FWHM's at different instrumental resolving powers.  The different coloured data points, linear trends 
and information in each panel represent the different resolving powers highlighted in each panel.  From top left to bottom right the FWHM's employed were 1.09, 2.00, 3.00 and 
4.00\AA\ respectively and are shown above each panel.  The plots highlight the increased scatter for the lowest resolution data and also we can see the decreasing correlation between 
the FEROS and MW $S$-indices with increasing bandpass width.}
\label{s_triangular}
\end{figure*}

To probe the origin of the effects mentioned above, and also to better study how to calibrate spectral $S$-indices onto the widely used MW system, which is important for 
studying the age-activity-rotation relation and for measuring the expected level of RV uncertainty for stars, we decided to artificially degrade our spectral resolution from our 
operating resolution down to below the working resolution of the Gray et al. data.  Figure~\ref{s_triangular} shows the changing calibrations onto the MW system and their associated 
rms scatter as a function of both changing resolving power and increasing FWHM of the HK core triangular bandpasses, where we show FWHM's of 1.09, 2.00, 3.00 and 4.00\AA. 

From these plots we can see two interesting features.  Firstly the rms scatter in each individual panel is fairly flat until we reach low resolving powers of around 2$'$500.  The scatter significantly 
increases below this value, which is close to the resolution of the \citet{gray06} NStars survey and can explain the higher level of scatter we see between our values and the Gray et al. 
data.  Secondly, when we move to larger bandpass widths, the slopes of the calibration relations significantly increase.  Using the 1.09\AA\ FWHM, which we recall is the width of the 
filter bandpass at MW, we see the linear trend gradient 
is almost parallel to the 1:1 relationship, highlighted by the dotted lines, for the higher resolution calibrations.  The gradient of the linear calibrations begin to steeply increase as we 
increase the bandpass width.  This increase in gradient, and hence 
departure from 1:1, shows the data become less correlated with the increase in core bandpass width,  however statistically the Pearson linear 
correlation coefficients are 0.98 or better until a width of 4\AA, and are still correlated at the level of 0.86 with a width of 6\AA.  We do see a small decrease in the correlation for the 
lower resolution data which changes from 0.99 at $\sim$46$'$000 resolving power down to 0.93 at $\sim$1$'$500 for a width of 1.09\AA, and for a width of 6\AA\ the change in 
the correlation 
is from 0.86 down to 0.76 at these resolving powers.  Note that we also include values for widths of 5.00 and 6.00\AA\ even though they are not shown in the figure, and 
report that once the bandpass FWHM reaches 6\AA\ then the rms around a straight line fit does not change with decreasing resolving power.  The increased rms with bandpass width 
comes about due to an increased influence from the calcium line shapes, representing regions not associated with the chromospheric plaques.

We do not show the same analysis as in figure~\ref{s_triangular} for square bandpasses at the cores as the fits are extremely similar, however we test square bandpasses given 
previous works have employed this methodology (e.g. \citealp{gray03}; 
\citealp{henry96}).  We find a similar rms scatter for the lower widths than the triangular method, however once the bandpass widths reach 4\AA\ it seems that square bandpasses 
produce a lower rms scatter around a linear fit for higher resolution data than using triangular bandpasses.  We note that there may be a need for polynomial fits to such wide square 
bandpasses and not a linear trend, but to confirm this many more stable calibration stars over the range are needed.

The changing correlation when the widths are increased follows a similar pattern to that of the triangular method.  It also appears that the data are more correlated 
when employing the square method with bandpass widths of 3\AA\ or greater.  Again more data are required to confirm this finding.  But this suggests that for lower resolution 
data sets, when large bandpasses are required to sample the line core, it is better to employ square and not triangular core filters.

We find that once the resolving power of the spectrograph reaches below ~2$'$500 the rms scatter from the linear calibrations onto the MW system of measurements significantly increases, 
yet no systematic is apparent.  This explains why the rms around the fit to the Henry et al. dataset is at the same level as the scatter to the Wright et al. data, even though 
Henry et al. observed the sample at much lower resolution.  We note there is a sharp drop in rms at around a resolving power of 2$'$500 and it is independent of the calibration method 
used i.e. 
altering the FWHM of the triangular bandpass or using a square bandpass gives rise to the same function.  The flat trend at higher resolving powers also tells us that MW $S$-index calibrations 
with resolving powers above 20$'$000 are as precise as those measured at much higher resolution. 

\section{Activity Distributions}

\begin{figure}
\vspace{5.5cm}
\hspace{-4.0cm}
\includegraphics{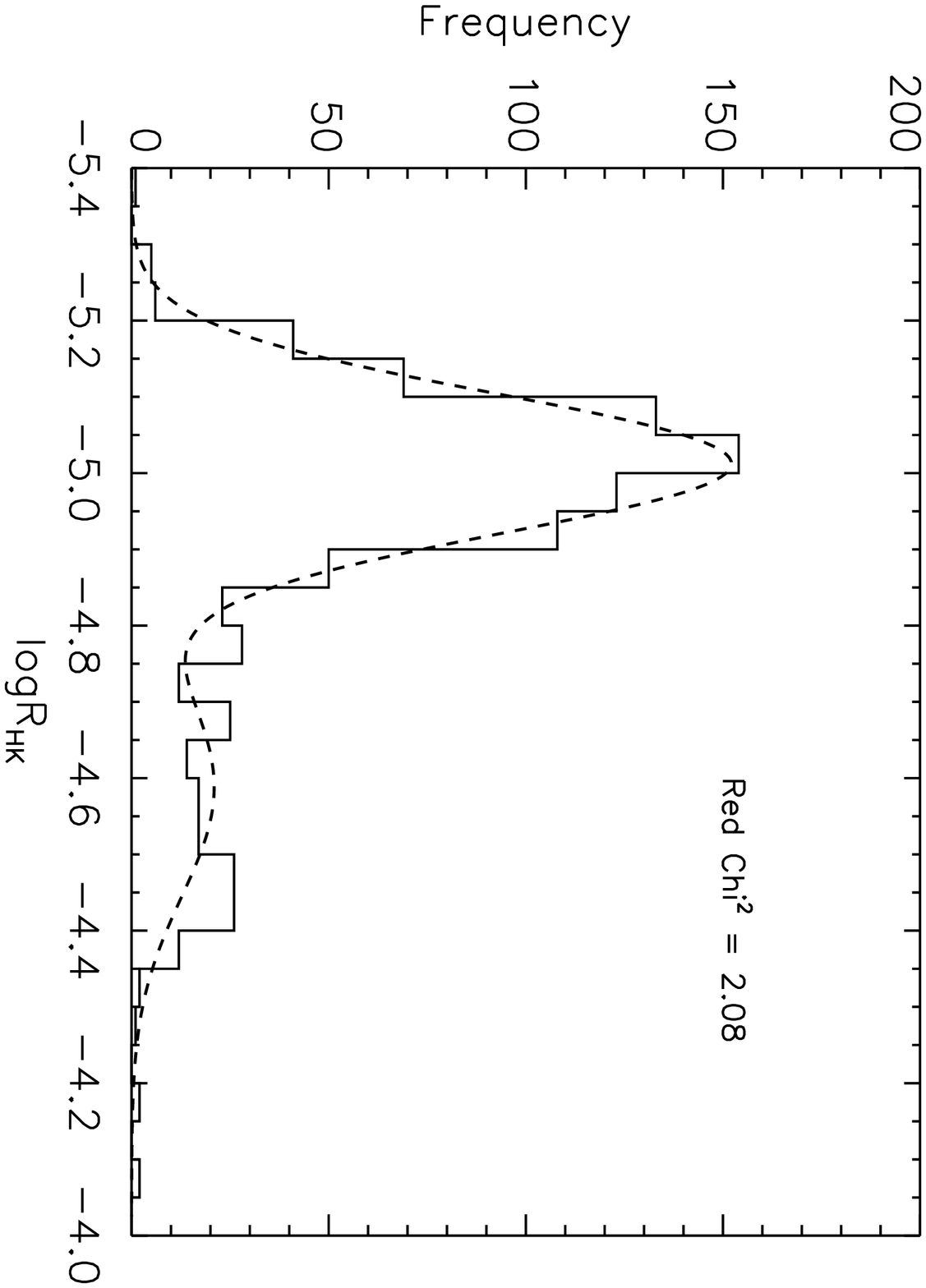}
\includegraphics{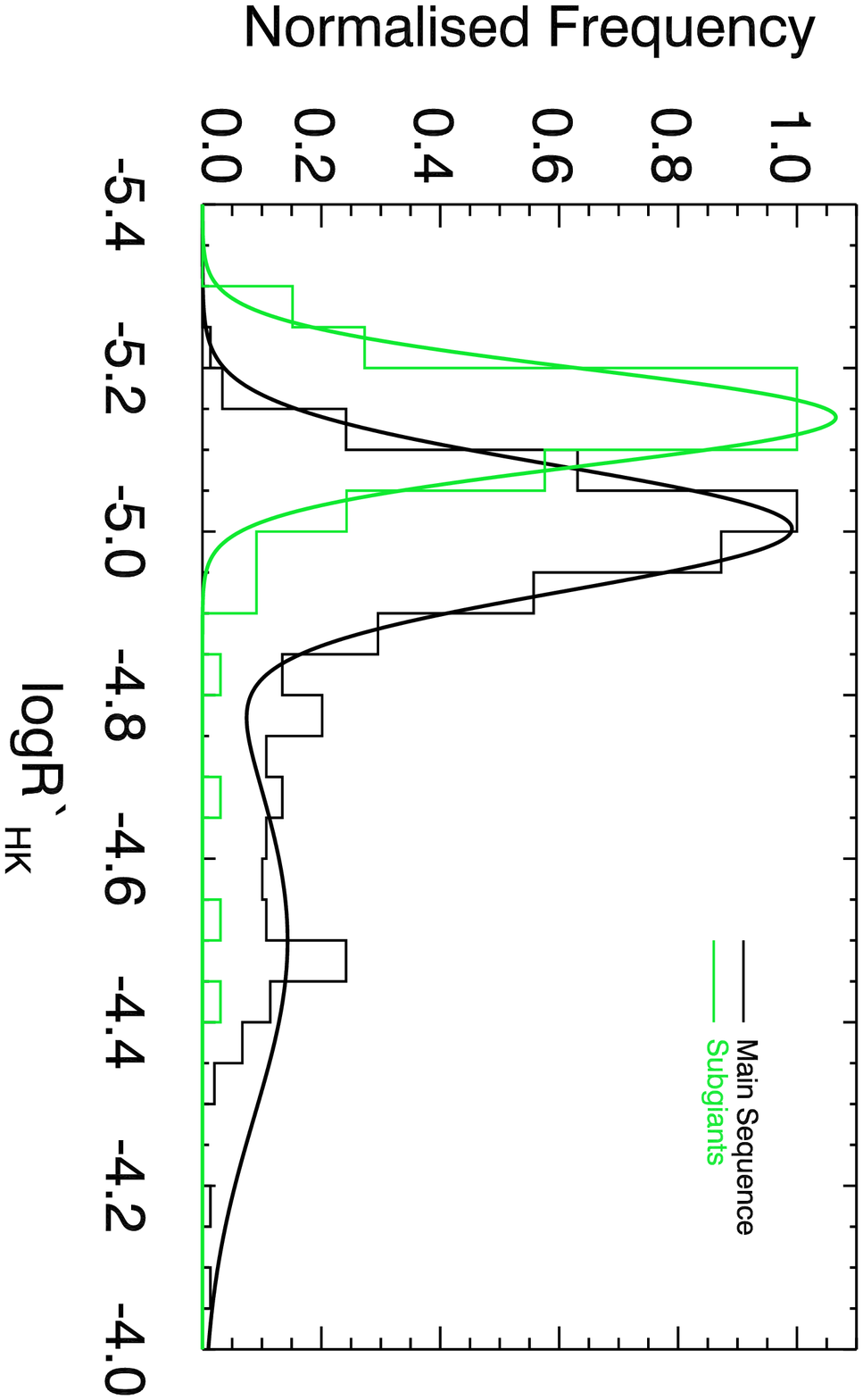}
\vspace{6.1cm}
\caption{The top panel shows the distribution of log$R'_{\rm{HK}}$ for our full sample of stars.  Most stars are located around the inactive part of the 
distribution ($\le$-4.8).  We  note there is a lack of stars around the Vaughan Preston gap and then a slight indication of the expected secondary active bump in 
this bimodal distribution.  We over plot the best fit bimodal function by the dashed curve and highlight the reduced $\chi$$^{2}$ of 2.08 we find.  The lower panel shows the 
sample split by main sequence stars (black) and subgiants (green).  The subgiants occupy a more inactive state than the main sequence stars.}
\label{rhk_dis_fit}
\end{figure}

In the top panel of Fig.~\ref{rhk_dis_fit} we show the overall chromospheric activity distribution for our sample of stars.  There is a clear over abundance of inactive stars 
(log$R'_{\rm{HK}}$~$\le$~-4.8~dex) 
that conforms to a Gaussian distribution.  Beyond a log$R'_{\rm{HK}}$ of -4.8 there is a dearth of stars and then a slight bump again of active stars after -4.5~dex.  Other works listed 
above have also confirmed this bimodal distribution and therefore we highlight the best fit bimodal model to the data by the dashed curve.  The reduced chi-squared ($\chi$$^{2}_{\nu}$) value 
is shown and found to be 2.08.  The largest discrepancy between the best fit and the data comes from the so-called $VP ~Gap$ region 
(-4.8~$\le$~log$R'_{\rm{HK}}$~$\le$~-4.6).  Given the earlier metallicity bias due to our sample stars being monitored for current and future planet searches (see Paper I), our selection 
method is biased toward inactive stars, hence we have a large number of stars that have settled in the inactive branch of the activity distribution.

The lower panel in Fig.~\ref{rhk_dis_fit} shows the distributions of activity if we split the sample into MS stars (black) and SG stars (green) using the prescription we mention above for what 
constitutes both classes of objects.  In fact we tested only selecting stars $\pm$0.45~magnitudes from the Hipparcos MS and the distribution was statistically similar to the MS distribution we show 
here.

\begin{figure}
\vspace{4.5cm}
\hspace{-4.0cm}
\includegraphics{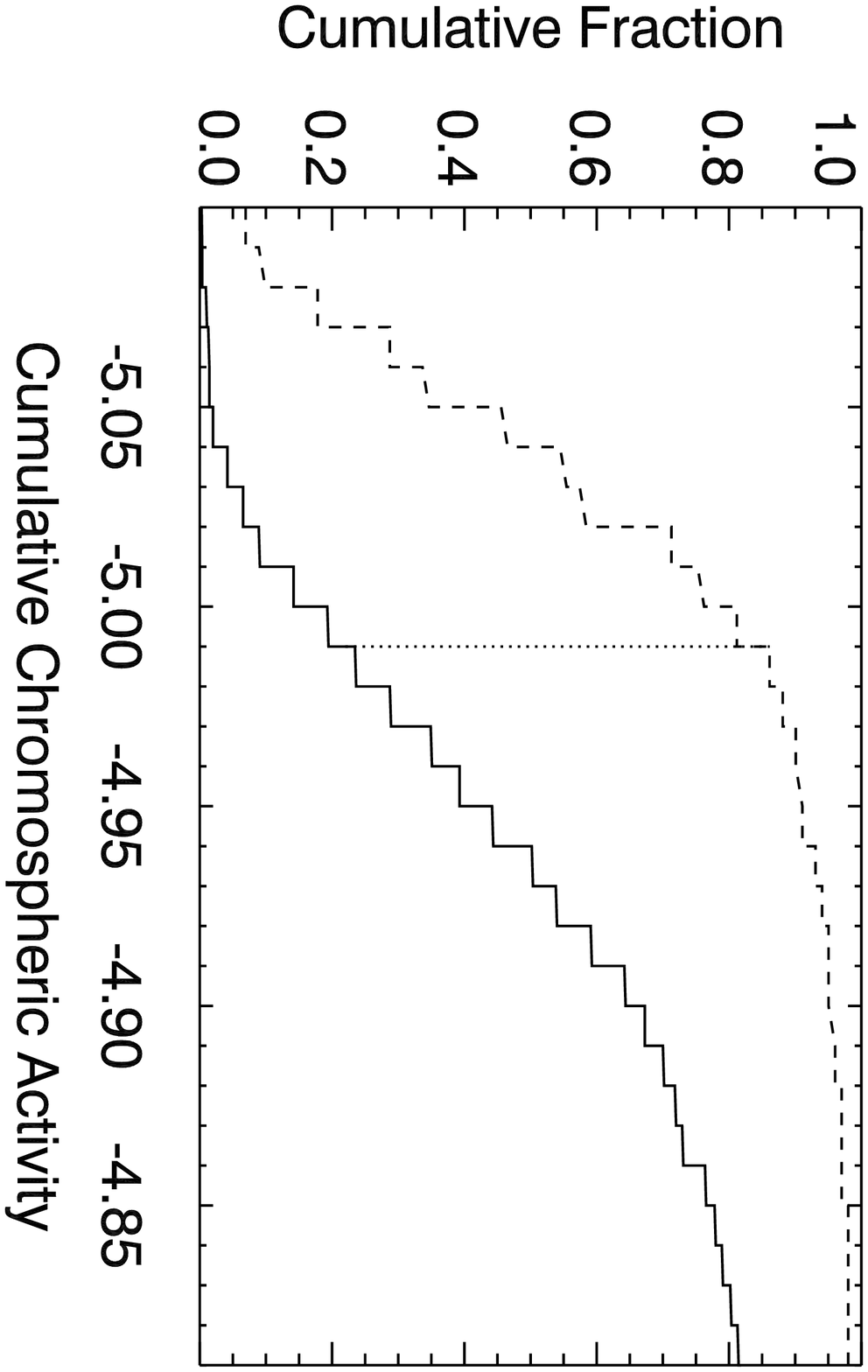}
\vspace{0.0cm}
\caption{The cumulative distributions for both main sequence (solid curve) and subgiant stars (dashed curve) as a function of the cumulative chromospheric activity.  The dotted 
vertical line represents the largest deviation between the distributions and hence relates to the KS test D statistic.  Clearly both these populations are not drawn from the same parent 
distribution at a high level of significance.}
\label{rhk_kstest}
\end{figure}

The distributions have inactive peaks of -5.004 and -5.140 for the MS and SG distributions respectively and the difference of -0.136~dex in the log$R'_{\rm{HK}}$ peaks highlight the 
mean time evolution of FGK stars when they traverse from the MS across to the SG branch.  A KS test yields a D-statistic of 0.664 (Fig.~\ref{rhk_kstest}), which relates to a very low 
probability ($<$10$^{-32}$\%) that these two populations are drawn from the same parent distribution.  Utilising the activity-age relation from \citet{mamajek08} this peak-to-peak difference 
relates to a time difference of +2.5~Gyrs.  We do note that there is a lack of good quality ages to calibrate the age-activity relationship for the older field population of stars, hence 
this value may have a fairly high statistical uncertainty.  We also draw your attention to the discussion in $S$5 to highlight the limits of this age-activity relation, in particular its 
applicability to evolved SG stars.  Given we have estimated masses for all objects by fitting evolutionary tracks, we know there is no significant mass 
offset between the MS and SG samples and hence the time evolution value we find here is not related simply to the faster evolution of higher mass stars.  This confirms that there is 
indeed a continuation of the age-activity relationship for stars with ages beyond the age of the Sun, even though it may be significantly reduced in comparison to pre-VP gap stars, and 
therefore more work is needed to better constrain the age-activity relationship for older field stars.

Another difference between the two populations appears to be the width of the gaussians after the VP gap.  The MS inactive sample has a gaussian sigma of 0.076, whereas the SG sample 
has a sigma of only 0.059, lower by 0.048 when differenced in quadrature.  This difference in width may indicate that when stars transition to the SG branch and their convective 
envelopes deepen then their activities become more stable and hence their cycle spans decrease.  This could be a direct consequence of the increased depth of the convective envelope 
since the magnetic fields may rise to the surface as buoyancy instability is less hindered, due to the lower density of the envelope material, and hence large changes in the magnetic 
dynamo and wind braking mechanism is thus reduced.  Studying the long term activity cycles of a large sample of SGs would allow a detailed look at any changes in the dynamo 
with changing depth of the convective envelope, like the A and I rotation period - activity cycle sequences (\citealp{bohm-vitense07}).  We do note that the SG sample spans a different age 
distribution than the MS sample but only by a Gyr or so, which argues in favour of a reduced braking mechanism i.e. the slower a star gets the more difficult it becomes for 
magnetic braking to slow the star's rotation even further. \rm

\begin{figure}
\vspace{5.5cm}
\hspace{-4.0cm}
\includegraphics{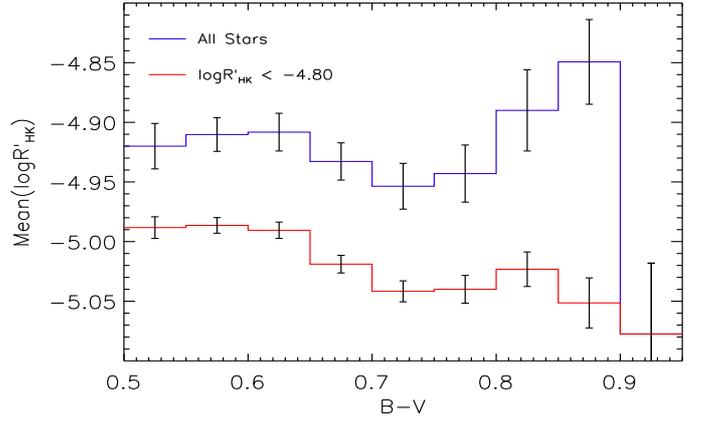}
\vspace{0.5cm}
\caption{Histograms of the mean log$R'_{\rm{HK}}$ as a function of the Hipparcos $B-V$ colour index for all the sample (blue) and only for the inactive stars (log$R'_{\rm{HK}}$~$\le$~-4.80~dex) 
located beyond the Vaughan-Preston gap (red).  Both samples are flat for the hotter stars, yet they diverge when approaching the cooler K stars.}
\label{rhk_col1}
\end{figure}

A more simple explanation for the increased width of the inactive star MS distribution could be through sample selection bias.  Fig.~\ref{rhk_col1} 
shows the mean distribution of log$R'_{\rm{HK}}$ across all $B-V$ colours in our range of interest for all dwarf stars in our sample (blue histogram) and for only the inactive stars that we, 
and other groups, generally focus on for planet search purposes (red histogram), which is more or less stars after the VP gap (log$R'_{\rm{HK}}$~$\le$~-4.80~dex).  
Lower mass and cooler K dwarfs have a longer MS lifetime than hotter G dwarf 
stars and hence the coolest K dwarfs will not yet have reached the SG branch.  This figure confirms that cooler stars are also more active in general and if these old K dwarfs 
are mostly populating the MS then they will mainly occupy the region close to the VP gap, giving the inactive distribution a wider appearance than the SG branch stars 
that are mostly composed of hotter G dwarfs.

To test this we decided to look at how the mean activity changes as a function of spectral type.  In both histograms in Fig.~\ref{rhk_col1} the hottest stars 
($B-V$~$\le$~0.65) tend towards a flat trend, then there appears a small drop in the mean activity of stars for later colour indices.  Towards the coolest dwarfs the two distributions diverge 
from each other, with the plot including all stars rising again towards the most active mean value of -4.85 and the inactive subset's distribution continuing on with a flat trend, within the 
uncertainties.  All uncertainties were taken as the standard deviation of the values within each 0.5~dex colour bin, divided by root-N, where N is the number of stars contained in that bin.  
This change for cool K dwarfs highlights the relatively higher fraction of active stars when moving down the spectral sequence.  Given this and the large spread of activities for K dwarfs, 
the VP gap may not be a fixed region but changes width with spectral type.  The gap appears to widen for the coolest stars when compared with the hotter objects, indicating that when 
moving down the spectral sequence the decay of activity with time is more laboured and therefore there is an increased spin-down timescale, similar to the theorised increased 
spin-down timescale for mid-to-late M stars (see \citealp{jenkins09b}), albeit with lower significance.  

\begin{table}
\small
\center
\caption{List of FEROS spectrum Binaries}
\label{tab:spec_binaries}
\begin{tabular}{ccc}
 & & \\
\hline 
\multicolumn{1}{c}{}  & \multicolumn{1}{c}{Binaries} & \multicolumn{1}{c}{}     \\ \hline

HIP2368           &  HIP33427    & HIP66118   \\
HIP4062           &  HIP39007    & HIP67112   \\
HIP6155           &  HIP45621    & HIP81179   \\
HIP7109           &  HIP51884    & HIP98373   \\
HIP10365         &  HIP53499    & HIP103883 \\
HIP14050         &  HIP54926    & HIP103941 \\
HIP15301         &  HIP55304   &  HIP106336 \\
HIP21329         &  HIP58132    & HIP115342 \\
HIP22064         &  HIP60391    & HIP115659 \\
HIP29772         &  HIP65403    & HIP117713 \\
HIP31623         &  HIP65548   &                     \\

\hline 
\end{tabular}
\medskip
\end{table}

\subsection{Spectroscopic Binaries}

\begin{figure}
\vspace{6.5cm}
\hspace{-4.0cm}
\includegraphics{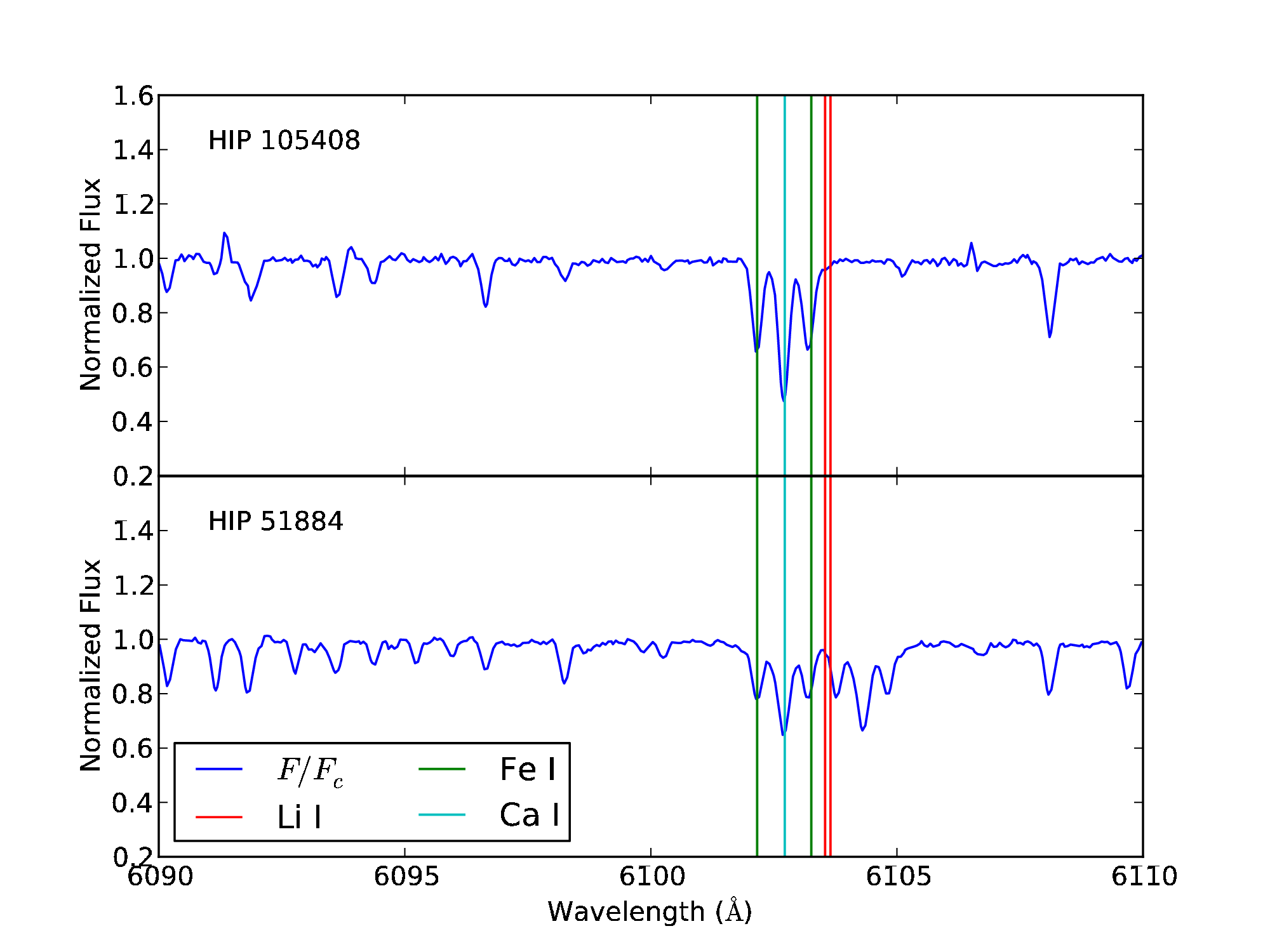}
\vspace{0.5cm}
\caption{The top panel shows a typical single stellar spectrum HIP105408 compared to the bottom panel which shows one of our detected spectroscopic binaries HIP51884, both 
represented by 
the blue curves.  The region we show is around the lithium line and we highlight the rest wavelengths of these lines in red, along with iron lines in green and a calcium line in light blue.}
\label{spec_binaries}
\end{figure}

In Table~\ref{tab:spec_binaries} we show our list of double-lined spectrum binaries that were discovered in our FEROS spectra, and Fig.~\ref{spec_binaries} shows a typical double-lined 
FEROS spectrum binary (HIP51884; bottom) compared to a typical single star (HIP105408; top).  Since we only have single epoch measurements for our 
sample we can not determine orbits for these binaries, yet they provide a good list of bright binaries that can be followed up in the future to further tighten the statistics drawn from the 
distributions of binary parameters.  Furthermore, with precision spectral deconvolution techniques, one can deconvolve these spectra into two single star spectra and use these as benchmarks 
for spectroscopic parameter testing.  For instance, if a clean enough deconvolution of the spectra can be made then measuring precise metallicities for each component can allow a test of 
such methods, given the stars should be coeval and hence underwent the same formation histories, assuming both components have not evolved long enough to undergo any dredge-up.  
Finally, we highlight these stars so that future planet search projects that focus on single solar-type stars will not use precious observing time on these stars.

\section{Rotation Velocities}

Given we have such a rich dataset, we initiated a method to extract precise rotational velocities for all our stars.  We follow a similar method to that employed by \citet{santos02}, whereby 
we measure the width of the cross-correlation function (CCF) that was used to measure RVs for our sample.  The CCF method is explained in more detail in \S~6 and we 
extract the CCF widths and use them to gain an accurate measurement for the stellar rotational velocity.  

\begin{figure}
\vspace{6.5cm}
\hspace{-4.0cm}
\includegraphics{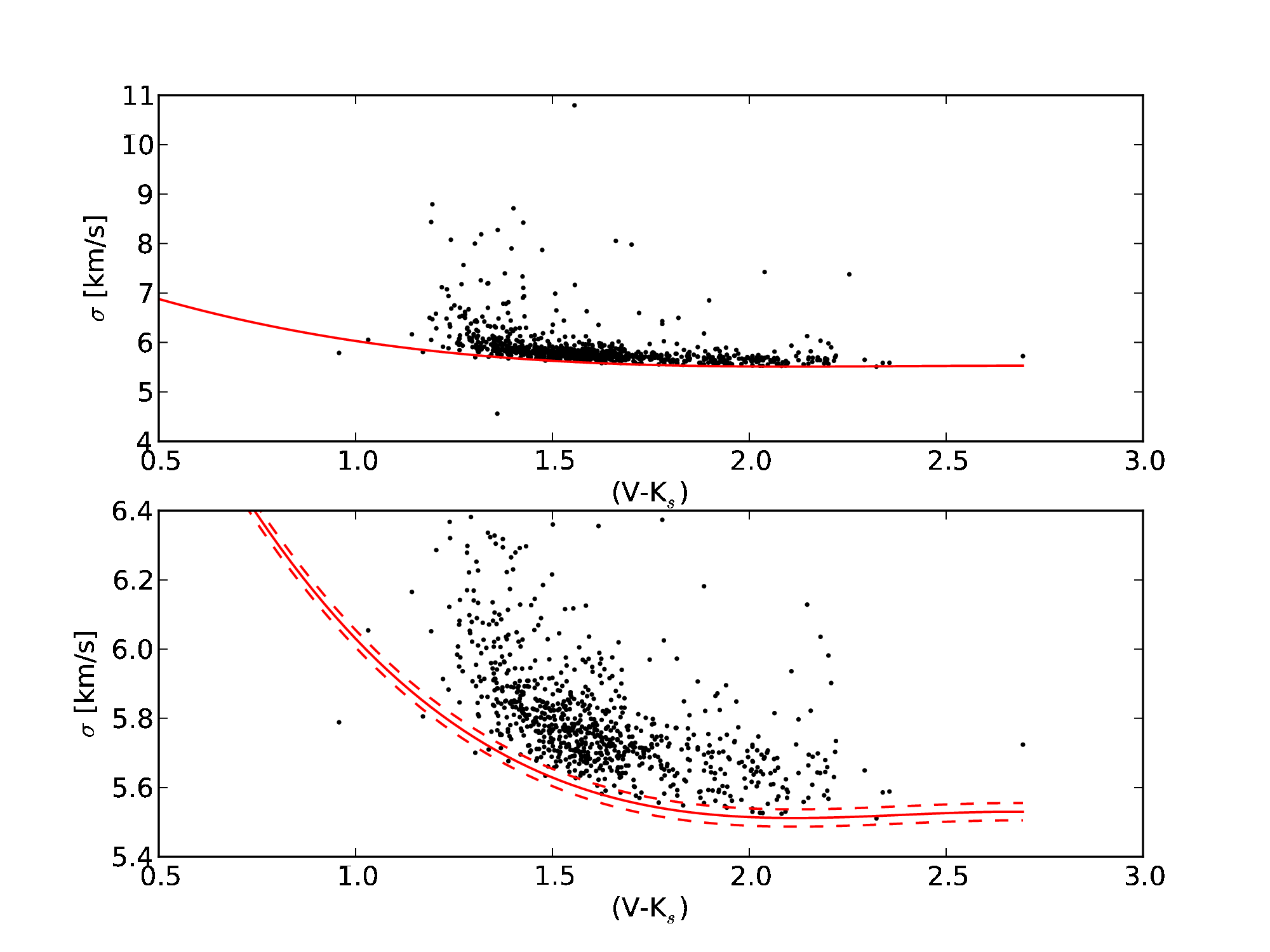}
\vspace{0.5cm}
\caption{The width of the cross-correlation function as a function of each star's $V-K_{s}$ colour index.  The top panel shows the full sample, including rapid rotators and 
the lower panel shows the zoomed region to highlight the slow rotators.  The red solid curves represent our best fit to the lower boundary for the slowest rotators, along with the 1$\sigma$ 
uncertainties shown by the dashed curves.}
\label{rot_vel}
\end{figure}

In Fig.~\ref{rot_vel} we show the widths of the CCFs for all our target stars as a function of their $V-K_{S}$ colours.  $V-K_{S}$ was used instead of $B-V$ since with the larger spectral 
energy distribution coverage and lower level of line blanketing the index is less affected by changes in metallicity and surface gravity and is therefore a better temperature indicator 
for FGK stars.  We can clearly see the trend of CCF width tailing off as we move to cooler stars.  This trend allows us to calculate a lower boundary ($\sigma_{\rm{o}}$) as a function of $V-K_{S}$, 
highlighted by the solid curve and also shown by the dashed curves are the 1$\sigma$ uncertainties on the 
fit.  The fit is a 3rd-order polynomial and is described by Eq$^{\rm{n}}$~\ref{eq:ccf_fit}.  Now we can simply take the width of each star's CCF ($\sigma$) and by application of 
Eq$^{\rm{n}}$~\ref{eq:vsini} we can calculate any star's $v$sin$i$.  The constant of proportionality in this equation ($A$) we set to be equal to 1.9$\pm$0.1, the value found for FEROS spectra by 
\citet{melo01}.

\begin{equation}
\label{eq:ccf_fit}
\sigma_{\rm{o}} = 8.337 - 3.632(V - K_{s}) + 1.540(V - K_{s})^{2} - 0.214(V - K_{s})^{3} .
\end{equation}

\begin{equation}
\label{eq:vsini}
v\rm{sin}\emph{i} = A\sqrt{\sigma^{2} - \sigma_{\rm{o}}^{2}} .
\end{equation}

\begin{figure}
\vspace{5.5cm}
\hspace{-4.0cm}
\includegraphics{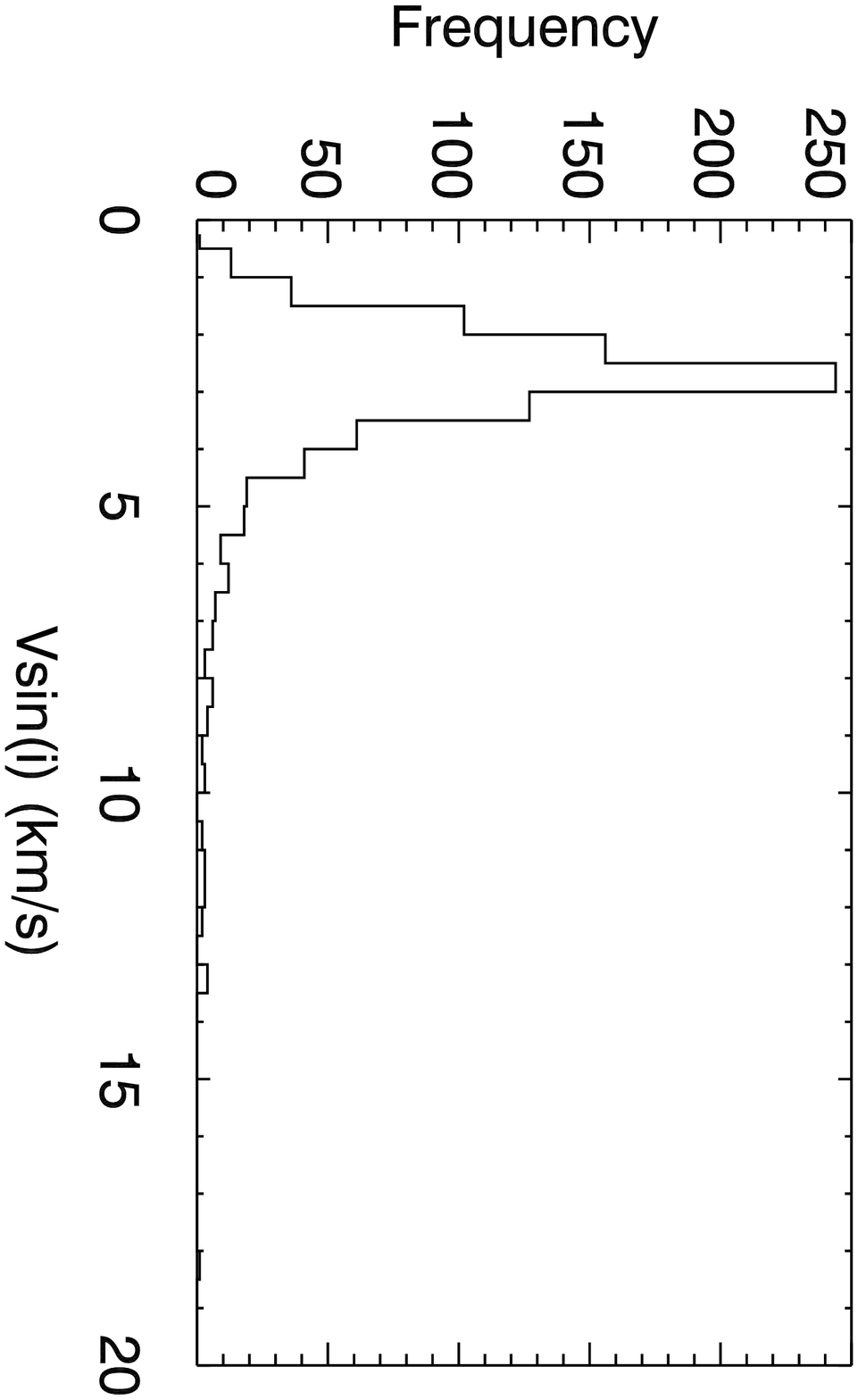}
\includegraphics{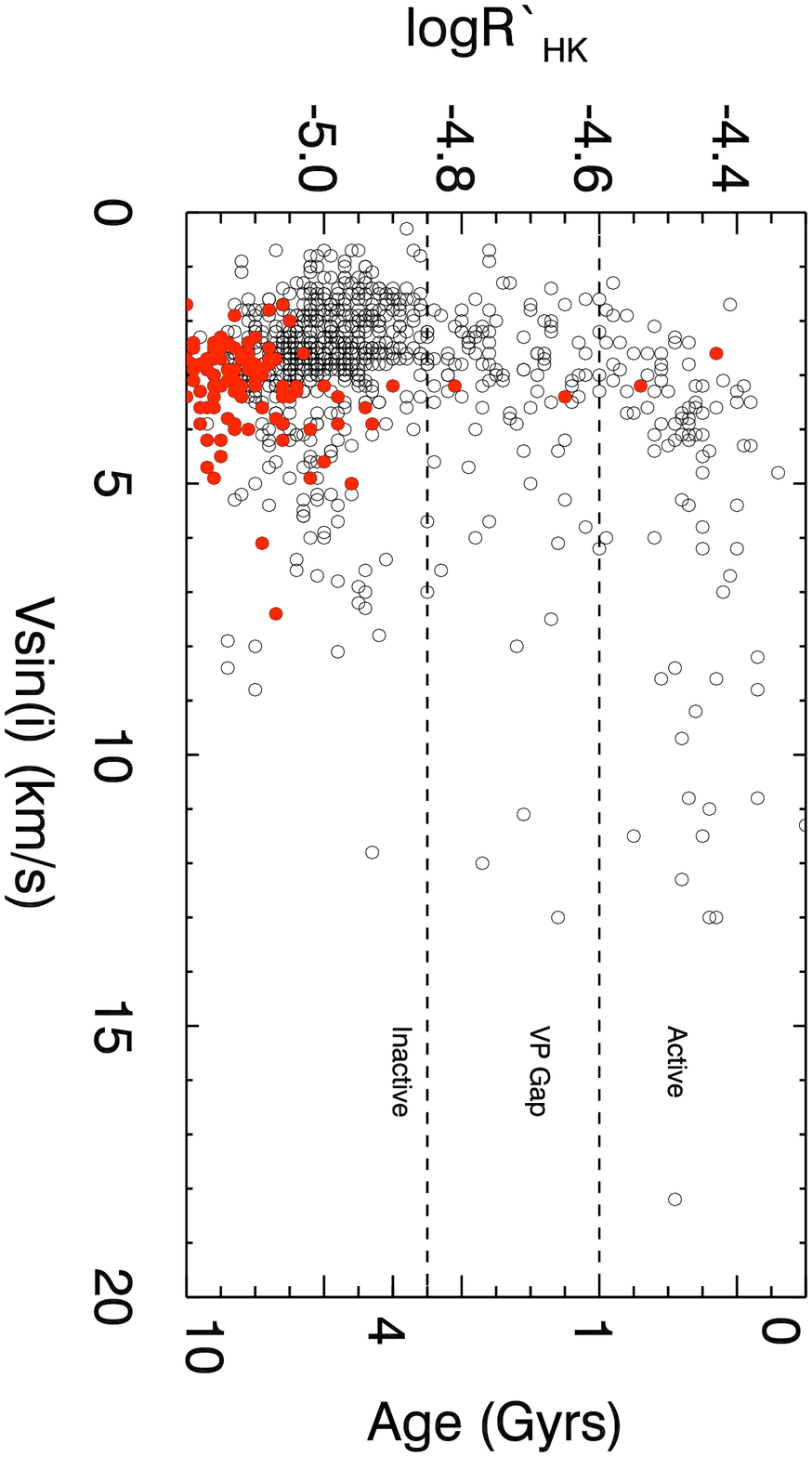}
\vspace{5.0cm}
\caption{The top panel shows the histogram of binned rotational velocities for the whole sample.  The sample peaks around 2.5~kms$^{-1}$ with a tail of faster rotating stars.  
The lower panel shows the chromospheric activities as a function of rotational velocity with the main sequence stars as open circles and subgiants are filled red circles.  Given these 
two quantities are correlated, due to the wind braking mechanism invoked to explain 
the activity distribution of such stars, we find an expected relationship whereby the stars that are deemed fast rotators are mostly found to have high log$R'_{\rm{HK}}$ activity indices.  The dashed 
lines bound the VP gap and the minimum measurable $v$sin$i$ is found to be 1.5kms$^{\rm{-1}}$.}
\label{act_rot}
\end{figure}

The top panel in Fig.~\ref{act_rot} shows the distribution of rotational velocities for our sample.  The distribution peaks between 2.5-3.0~kms$^{\rm{-1}}$.  
Given that Fig.~\ref{feros_hr} shows our sample to be mainly drawn 
from stars on or evolved beyond the MS, we expect our sample to peak at very low rotational velocities due to magnetic braking slowing the stars down throughout their main 
sequence lifetime.  The width of the distribution likely reflects the age scatter of stars, the initial $birth$ rotational velocities, the efficiency of the braking mechanism as a function of 
spectral type and probably most importantly the inclination angle of each system.  The tail of fast rotators are probably a subset of young stars that have recently moved onto the MS.

The lower panel in this figure shows the measured log$R'_{\rm{HK}}$ chromospheric activities for each star as a function of projected rotational velocity for MS (open circles) and 
SG stars (filled red circles).  These quantities have long been 
known to be correlated (\citealp{noyes84a}) since the magnetic braking mechanism is powered by the internal rotational dynamo of the star and hence as the star ages, the powerful magnetic 
fields interact with the stellar wind, carrying away angular momentum, slowing the star, which weakens the magnetic fields and decreases the stars magnetic activity (\citealp{kraft67}; 
\citealp{montesinos01}).  We can see the general trend here, albeit with a high degree of scatter due in part to the intrinsic variability of a star's chromospheric activity level.  As expected 
the majority of evolved SG stars are located in the bottom right of the plot with very low activity and low rotational velocity.  We also highlight 
the extremes of the VP gap by dashed horizontal lines which delimit our active and inactive stars.  On the right axis we also plot 
the mean age as a function of activity, determined here and in all proceeding plots that show age, using the latest incarnation of the age-activity relationship in \citet{mamajek08}.  Clearly 
the bulk of this sample consists of slowly rotating, old and inactive stars.  We note a caveat here in that although our sample range of $B-V$ colours are within limits of this relation, our 
SG star sample is not, as these stars are located at least one magnitude above the MS.  Therefore, we have extrapolated this relationship beyond the MS onto the SG branch.  Although, 
the convective envelopes of MS stars deepen as they move onto the SG branch we still believe an age-activity relationship holds true, even though using this relationship may not be 
strictly correct.  This point should be remembered in all future discussion comparing ages of SG stars against MS stars. 

Our sample consists mostly of late F to early K stars and it is necessary to briefly discuss the implications of this effective temperature range on the activity-rotation relation.  
Since the chromospheric activity is well correlated with the Rossby number (R$_{\rm{o}}$; \citealp{noyes84a}) we can look at the difference for a given rotational period that this span in temperature will 
produce in the activity-rotation relationship.  For the range of spectral types in our sample, R$_{\rm{o}}$ is around double the value for the late F's compared with the early K's.  For slowly 
rotating stars this relates to a difference of $\sim$0.7~dex in log$R'_{\rm{HK}}$, whereby the K dwarfs are more active by this amount.
All rotational velocities are shown in column 6 of Table~4. \rm

\section{Radial-Velocities}

\begin{figure}
\vspace{6.0cm}
\hspace{-4.0cm}
\includegraphics{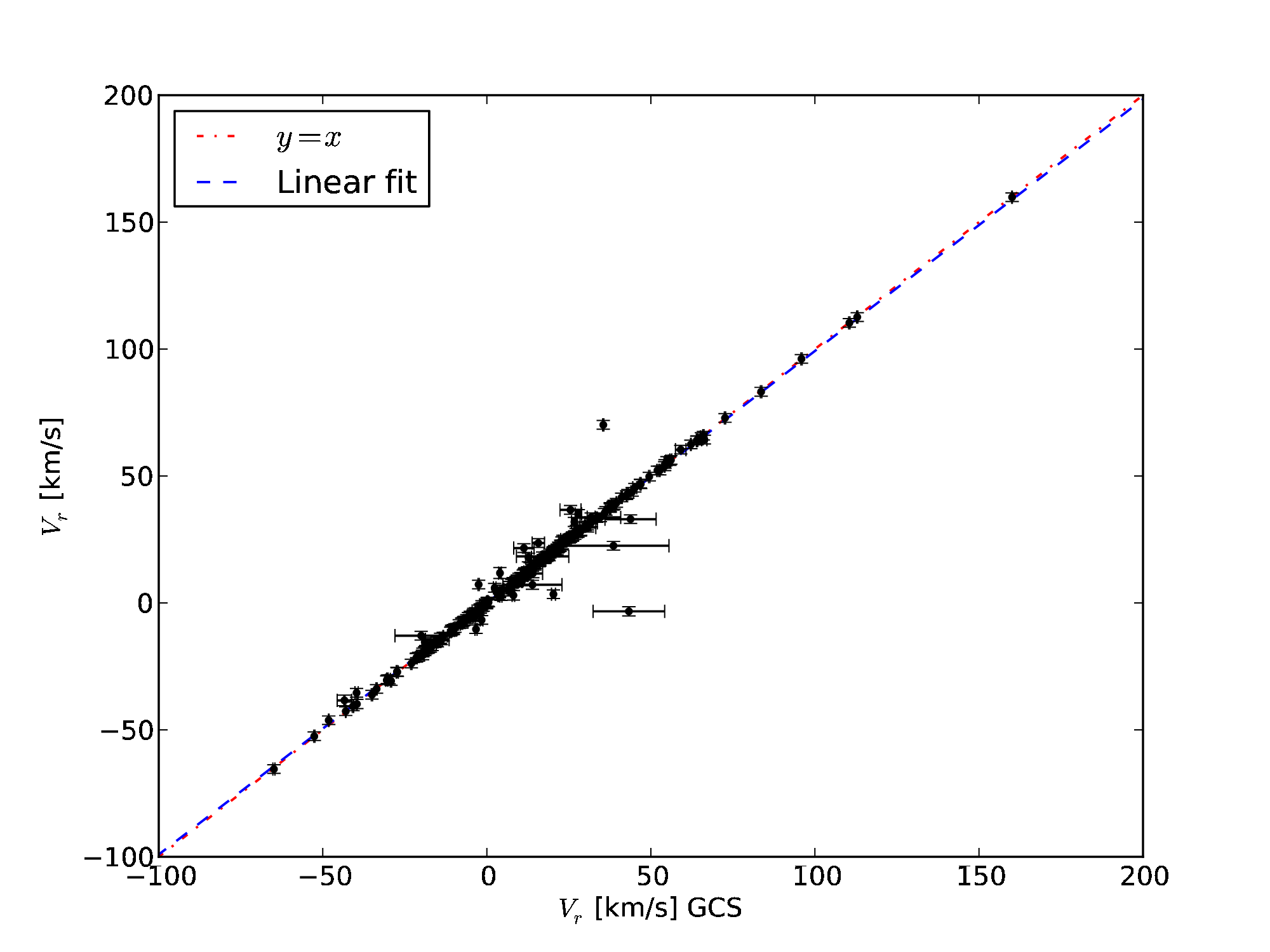}
\vspace{0.5cm}
\caption{A comparison of our derived FEROS RVs with those measured by the GCS.  The dot-dashed line represents the 1:1 relationship and the long dashed 
line is the best fit to the data.  The two are indistinguishable highlighting our robust values.}
\label{rv_gcs}
\end{figure}

FEROS is known to be stable over a fairly long baseline at better than the $\sim$10ms$^{\rm{-1}}$ level (\citealp{setiawan07}) and since we observed each star using the simultaneous 
ThAr mode we are able to 
measure accurate RVs that allow us to determine the full kinematic space velocity for each star.  In fact we found that the scatter in ThAr lamps themselves did not exceed 
300ms$^{\rm{-1}}$ for our entire sampling baseline of three years, highlighting the internal stability of FEROS.  We performed the cross-correlation method to obtain the absolute 
RV by cross-correlating against stars we had observed that overlapped with the GCS survey and which were not found to be binaries since binary motion could add large kms$^{\rm{-1}}$ scatter to our 
calibration.  The list of stars we used are shown in Table.~\ref{rv_templates} along with their absolute GCS measured RVs.

\begin{equation}
\label{eq:rv}
V_{r} = V_{r, stars} - V_{r, lamps} + V_{r, GCS}.
\end{equation}

\begin{equation}
\label{eq:mean_rv}
V_{r, final} = \frac{1}{N_{templates}} \displaystyle\sum\limits_{i=1}^{N_{templates}} V_{r,i}.
\end{equation}

The calibrator stars were all observed with high S/N and were used as templates, such that we cross-correlated against each of the targets, measured the overall internal FEROS drift between 
each template observation and each star by cross-correlating the ThAr orders and subtracted this from the relative velocity.  Finally we added the GCS measured absolute velocity to get 17 
independent measurements of the RV for each star for which we determined their mean.  Eq$^{\rm{n}}$s~\ref{eq:rv}~\&~\ref{eq:mean_rv} mathematically describe this procedure where $V_{r, stars}$ 
is the relative measured RV, $V_{r, lamps}$ is the relative drift correction, $V_{r, GCS}$ is the measured GCS template RV, $V_{r, final}$ is the final mean 
RV and $N_{templates}$ is the number of template stars used in the calculation of the mean.  Our uncertainties were derived from the width of the CCF 
(\citealp{tonry79}) and were then propagated through to the mean.  The vast majority of uncertainties in our sample were between 1.6-1.8kms$^{\rm{-1}}$, with a very small tail of higher values.

\begin{table}
\small
\center
\caption{RV template stars}
\label{rv_templates}
\begin{tabular}{cc}
 & \\
\hline 
\multicolumn{1}{c}{Star}  & \multicolumn{1}{c}{RV kms$^{\rm{-1}}$ (GCS)} \\ \hline

HIP6125          &  31.8$\pm$2.7    \\
HIP7693          &  26.0$\pm$0.3    \\
HIP11514         &  18.3$\pm$0.2    \\
HIP13889         &  11.5$\pm$0.2    \\
HIP13908         &  7.6$\pm$0.1    \\
HIP14774         &  21.4$\pm$0.4    \\
HIP79149         &  -52.6$\pm$0.2    \\
HIP81229         &  -17.8$\pm$0.1    \\
HIP88650         &  14.6$\pm$0.5    \\
HIP90896         &  11.6$\pm$0.1    \\
HIP96881         &  -10.8$\pm$0.2    \\
HIP98599         &  -15.1$\pm$0.1   \\
HIP100359        &  12.4$\pm$0.3    \\
HIP100474        &  -1.5$\pm$0.1    \\
HIP100649        &  10.4$\pm$0.1    \\
HIP114590        &  4.3$\pm$0.3    \\
HIP114967        &  7.8$\pm$0.2    \\

\hline 
\end{tabular}
\medskip
\end{table}

Fig.~\ref{rv_gcs} shows all our 235 overlapping stars against the GCS, including the templates.  We find a 1:1 best fit marked by the dot-dashed line.  
The true 1:1 relationship is represented by the long dashed line and since these two lines overlap to a high degree this highlights our method is robust with only a few outliers ($<$5\%) 
present in the plot, which are mostly stars with large uncertainties from low S/N and/or fairly high rotational velocities.  

\begin{figure}
\vspace{5.5cm}
\hspace{-4.0cm}
\includegraphics{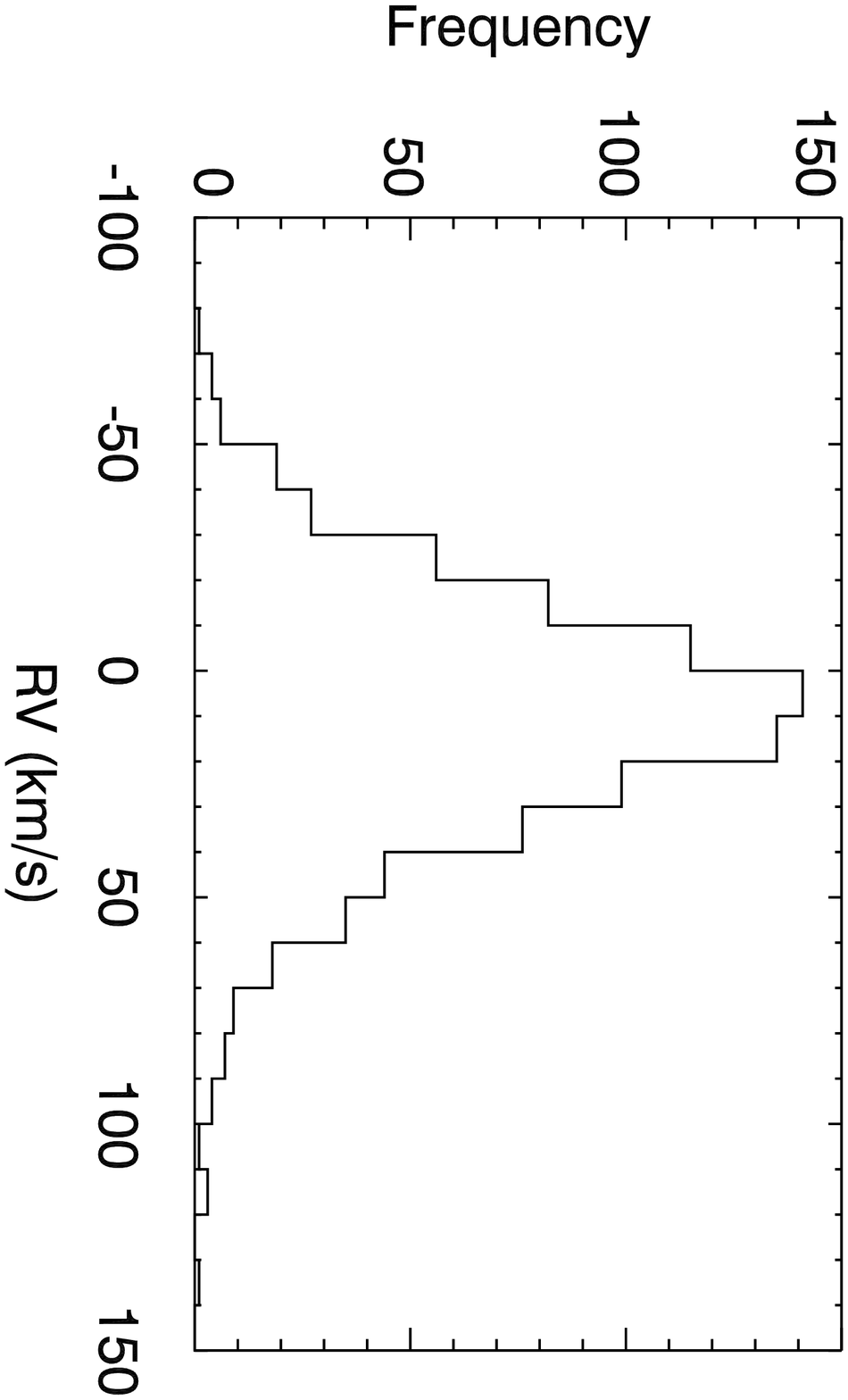}
\includegraphics{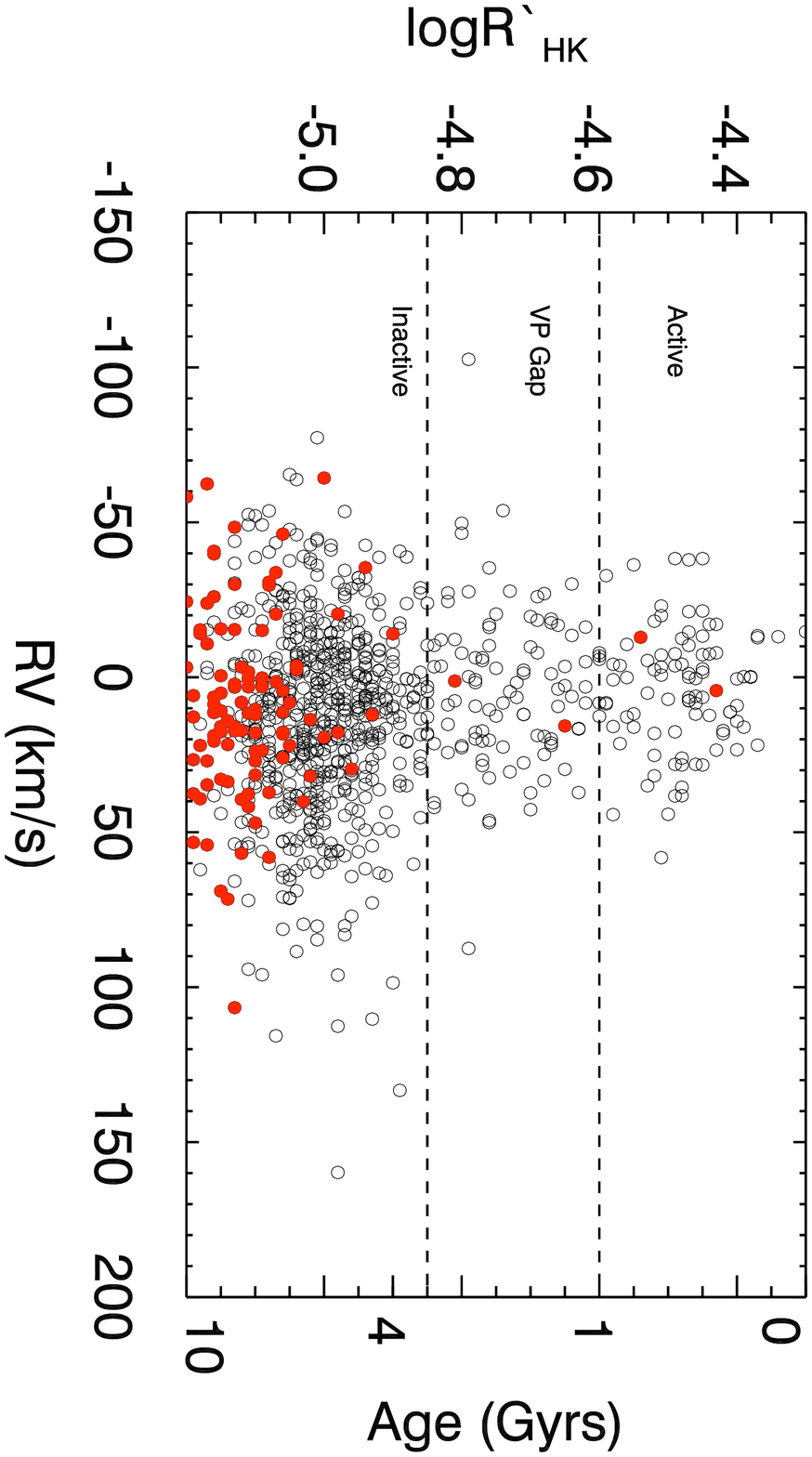}
\vspace{5.0cm}
\caption{The top panel shows the histogram of binned RVs for the sample.  The sample peaks around 5~kms$^{\rm{-1}}$ and follows a gaussian distribution.  
The lower panel shows the chromospheric activities as a function of RV and age on the right hand $y$-axis.  Open circles are main sequence stars and filled red circles are 
the subgiant stars.  Dashed lines bound the active, inactive and Vaughan Preston 
gap stars.  We find an increase in the dispersion over the gap region of $\sim$10kms$^{\rm{-1}}$.}
\label{act_rv}
\end{figure}

The distribution of our RVs are shown in the top panel of Fig.~\ref{act_rv}.  As expected from other large samples like the GCS we see a well formed Gaussian distribution 
with a mean around 10kms$^{\rm{-1}}$.  The lower panel shows the measured activities plotted as a function of RV.  Again the VP gap region is highlighted, along with active and inactive 
stars.  The MS and SG stars are shown by black and red filled circles respectively.  The distribution shows a column of stars spread around the mean of the overall distribution and then 
after the VP gap at the inactive side, where we see a slightly larger spread in the RVs.  This increase is at the level of $\sim$10kms$^{\rm{-1}}$.  It also appears there is a small overabundance 
of stars clustered around a velocity of $\sim$100kms$^{\rm{-1}}$, which could be a small selectional bias.  We note that the MS and SG stars follow the same distribution in RVs.  
The radial-velocities are shown in column 7 of Table~4.

\section{Kinematics}

The kinematic space motions for our stars are derived using a right-handed coordinate system and making use of our measured RVs, the Hipparcos parallaxes and proper 
motions.  The U kinematic component is positive in the direction of the galactic center, the V component is positive in the direction of galactic rotation and W is positive towards the north 
galactic pole.  The galactic reference system was defined in the equinox B1950.0, so the accurate way of obtaining U,V and W is to precess the equatorial coordinates, proper motions and 
RVs to that equinox and then precess the result to the J2000.0 system. As an approximation we computed the galactic velocities using the values in the J2000.0 equinox and due 
to the observational uncertainties we are confident that our values are precise to better than 5kms$^{\rm{-1}}$ despite this approximation.  We also employ a correction to the local standard of rest given 
in \citet{dehnen98}.

\begin{figure*}
\vspace{9.0cm}
\hspace{-4.0cm}
\includegraphics{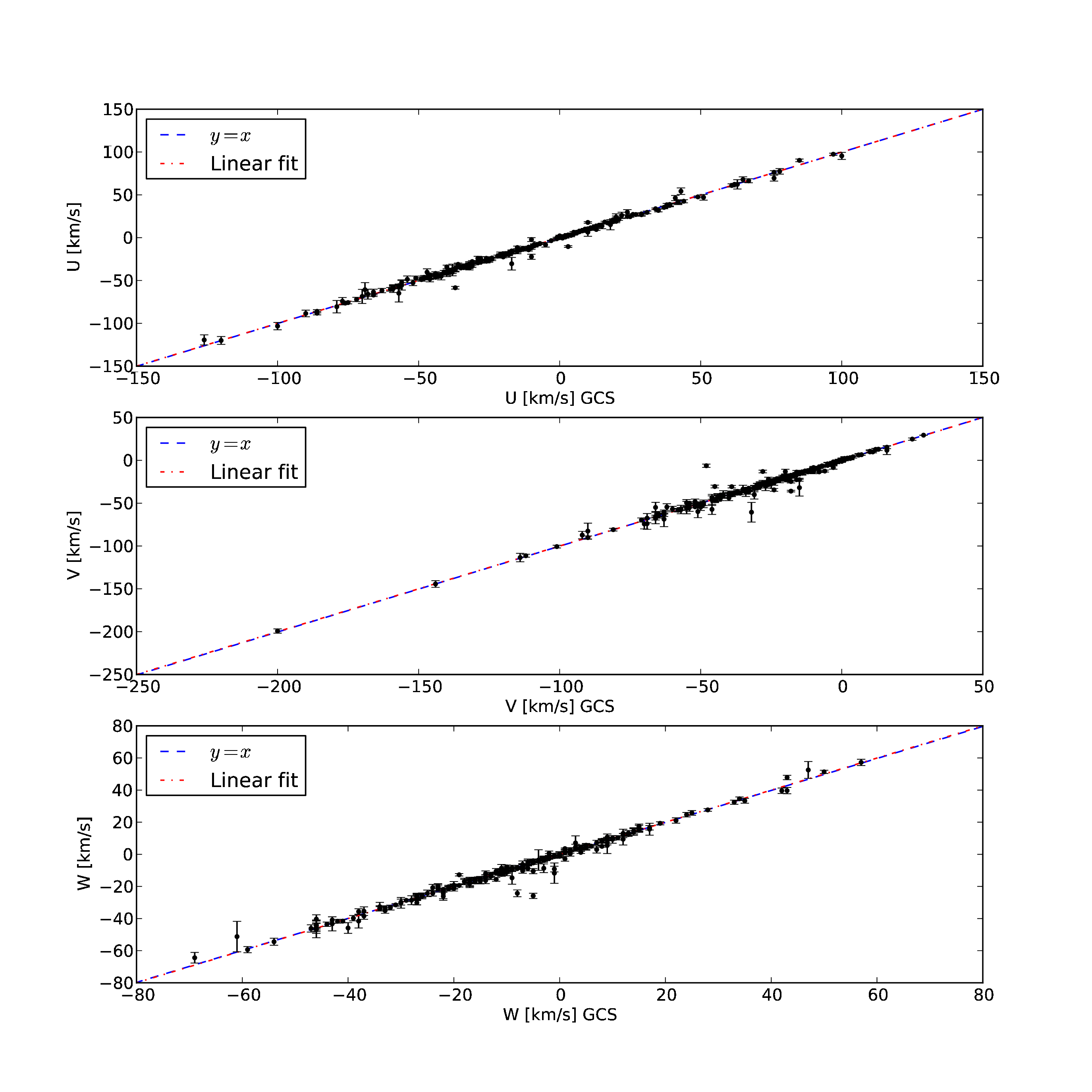}
\vspace{0.5cm}
\caption{A comparison between our measured UVW kinematic velocities and those of the GCS, from top to bottom respectively.  All three also show the 1:1 relationship (red dot-dashed 
line) and the best fit to the data (blue dashed line).  All three agree with those measured by the GCS with only a few outliers.}
\label{kinematic_comparisons}
\end{figure*}

Fig.~\ref{kinematic_comparisons} shows a comparison between the space velocities U , V and W obtained in this work and the values published in \citet{holmberg07} for 231 stars that are 
common to our sample.  We can see that the velocities are in good agreement with only a few outliers.  The standard 
deviation of the residuals is 3.0, 4.6 and 2.6 kms$^{\rm{-1}}$ for U , V and W respectively.  As mentioned above in $\S$6 the RVs are highly precise which is reflected in 
the low dispersion for the U component.  The higher dispersion in V can not be explained by errors in the distances because we should have seen a higher dispersion in W, which is not the 
case, so the increased scatter may come from the use of coordinates in the J2000.0 system.  All kinematic velocities are shown in columns 8-10 of Table~4.

\begin{figure*}
\vspace{4.5cm}
\hspace{-4.0cm}
\includegraphics{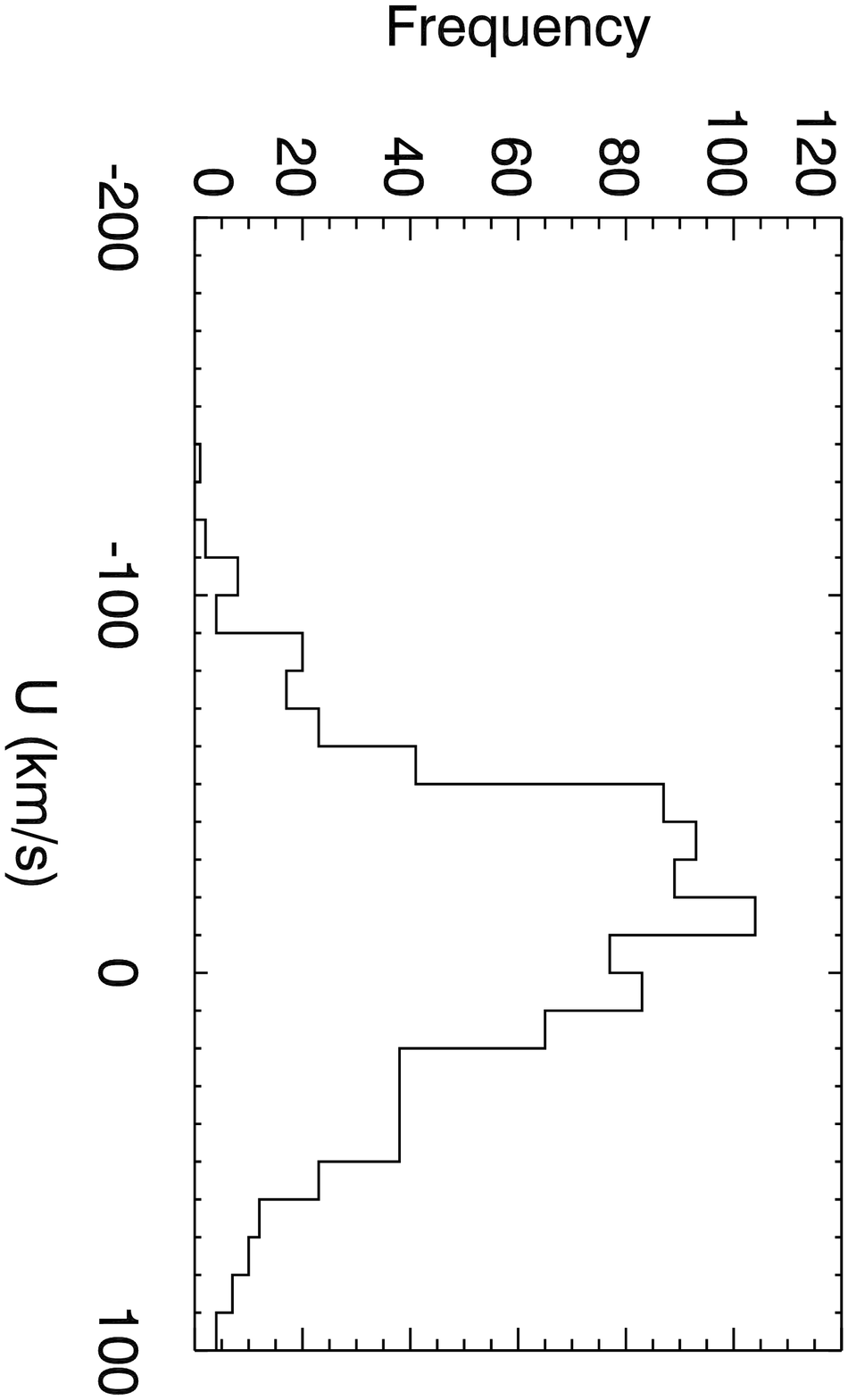}
\includegraphics{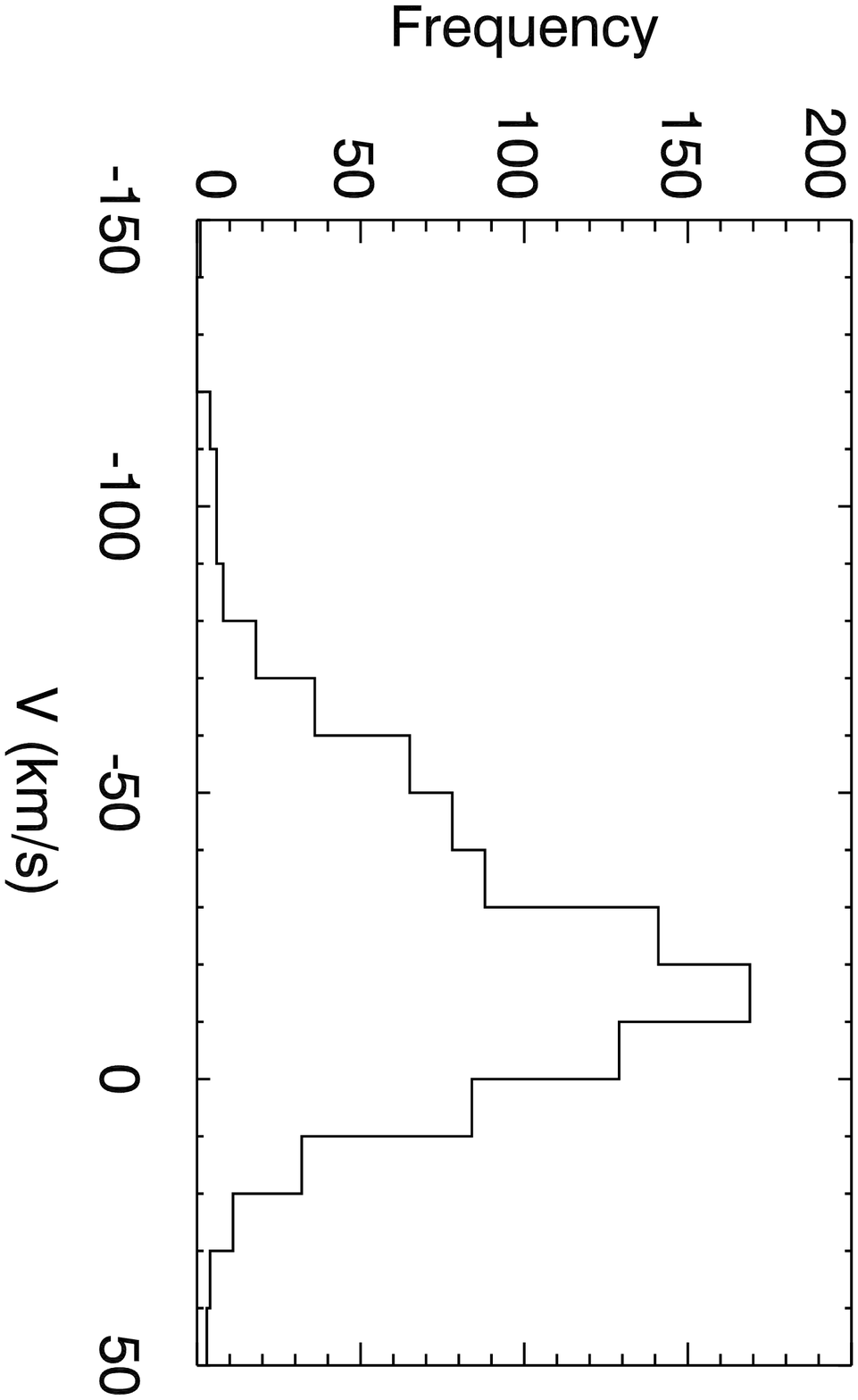}
\includegraphics{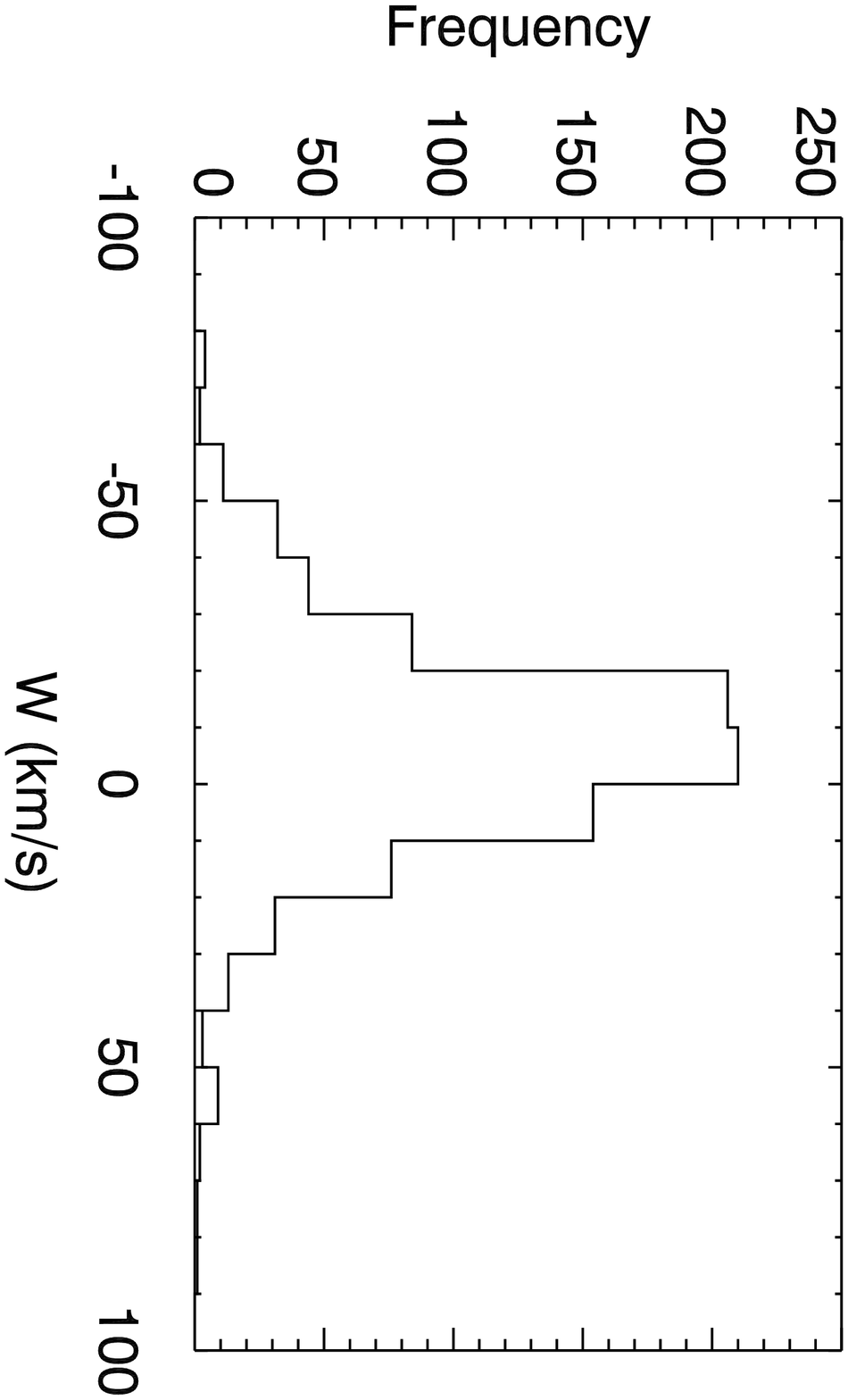}
\includegraphics{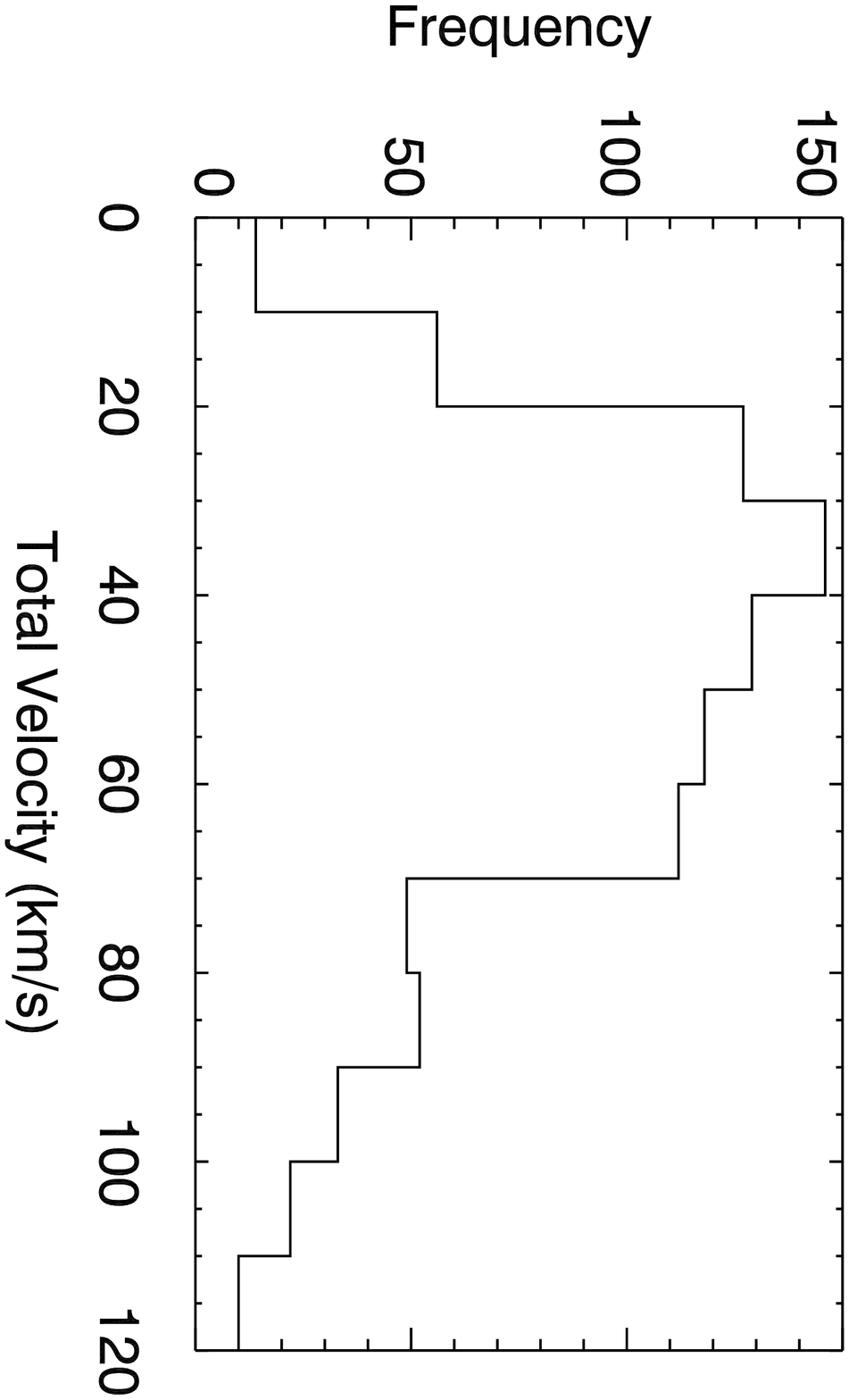}
\vspace{5.5cm}
\caption{The kinematic velocity distributions for our sample.  The U (top left) and W (bottom left) velocity distributions appear symmetrical around peaks just off zero.  
Whereas the V (upper right) and total kinematic velocity distribution (lower right) are far from symmetrical.  The V distribution shows a fairly high fraction of high proper 
motion objects moving against the plane of galactic rotation.}
\label{kin_dis}
\end{figure*}

The distribution of kinematic velocities can tell a lot about a sample of stars and hence we show our kinematic velocity distributions in Fig.~\ref{kin_dis} with U, V, W and total kinematic velocity ($\sqrt{U^{2}+V^{2}+W^{2}}$) represented.  The upper left panel shows the kinematic U 
component binned into steps of 10~kms$^{\rm{-1}}$.  The distribution generally follows a broad Gaussian profile or a sinc function, but given that this is a biased sample in general, the overall 
shape does not fully follow the true distribution of dwarf and SG stars in the solar neighbourhood drawn from a kinematically unbiased sample (see \citealp{holmberg07}).  However, 
we see that it does broadly conform to a distribution that has been drawn from a Schwartzchild distribution function (SDF).  The W component (lower left) also 
appears to follow a SDF having a fairly Gaussian shape.  However, the V component (upper right) appears to have a shallow/slow rise to a peak and then a fast drop off at greater than -10kms$^{\rm{-1}}$.  
The skewness in this distribution, which clearly is not SDF-like, comes from the velocity dependence of V on our sample of stars, as a function of galactocentric distance.  The 
density of disk stars increase exponentially towards the galactic center, i.e. inside the Sun's orbital circle around the galaxy, and the velocity dispersion of stars also increases towards 
the center of the galaxy, meaning we expect a higher fraction of stars to visit the solar neighbourhood from smaller galactic radii.  A phenomenon known as the \emph{asymmetric drift} tells us 
that the mean rotational velocity of a group of stars will lag behind the local standard of rest more and more with increasing velocity dispersion, or random motion within the population.  
Therefore, we would expect to find an overabundance of stars with lower V-type kinematic motions (see \citealp{binney00b}).  Finally the total kinematic 
velocity distribution (lower right) has a sharp rise at low velocities followed by a general shallower drop off towards the highest kinematic velocities.
 
\begin{figure*}
\vspace{3.5cm}
\hspace{-4.0cm}
\includegraphics{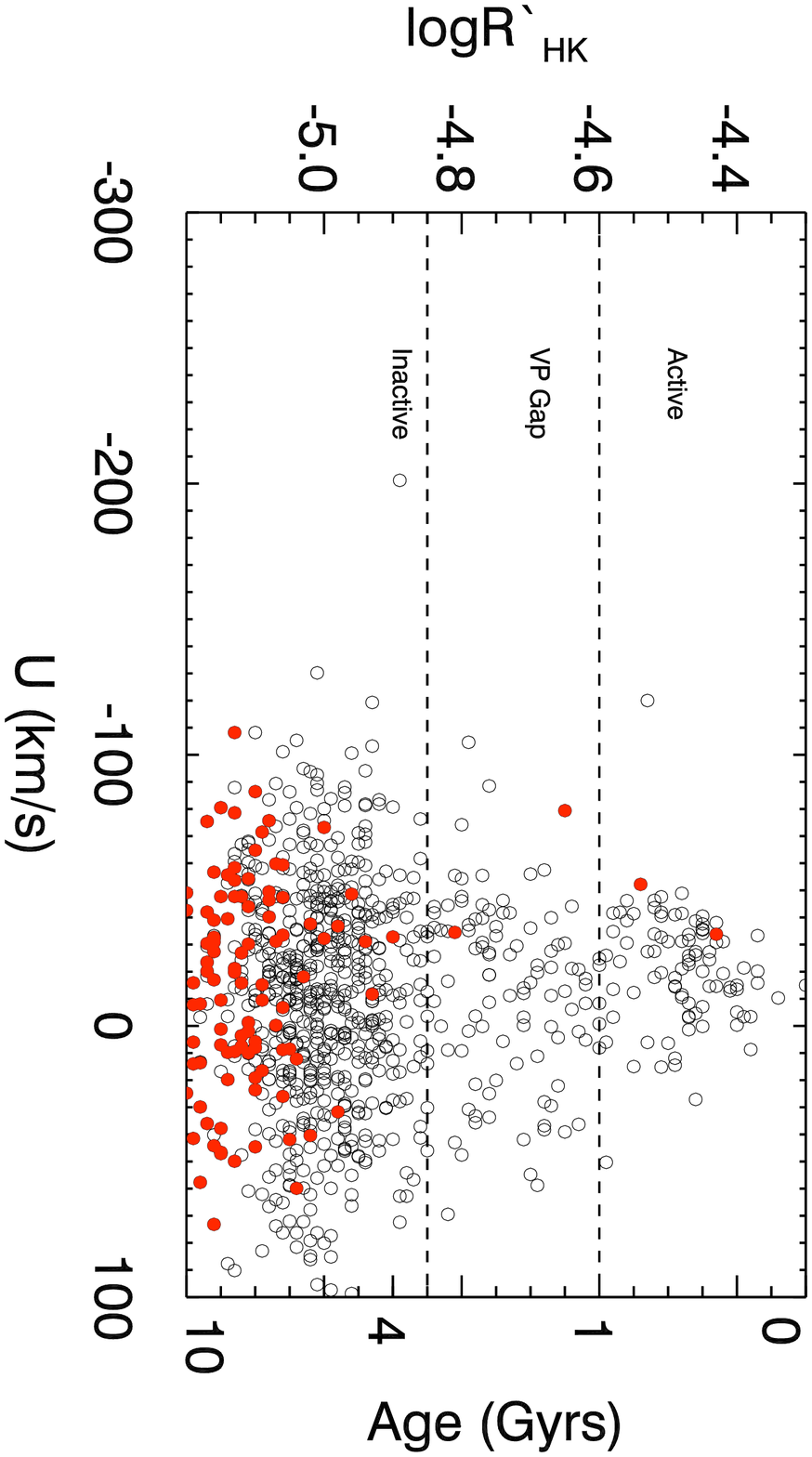}
\includegraphics{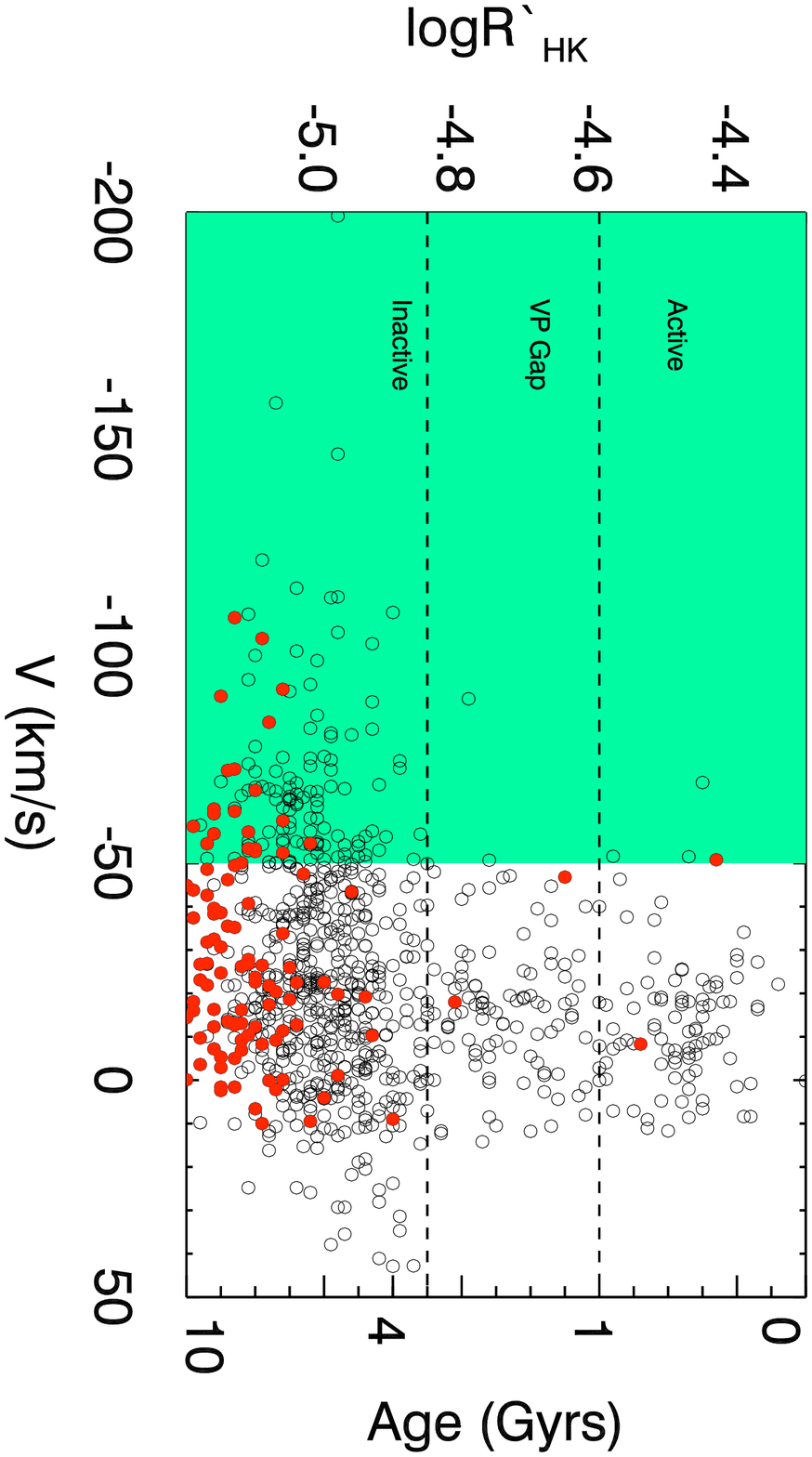}
\includegraphics{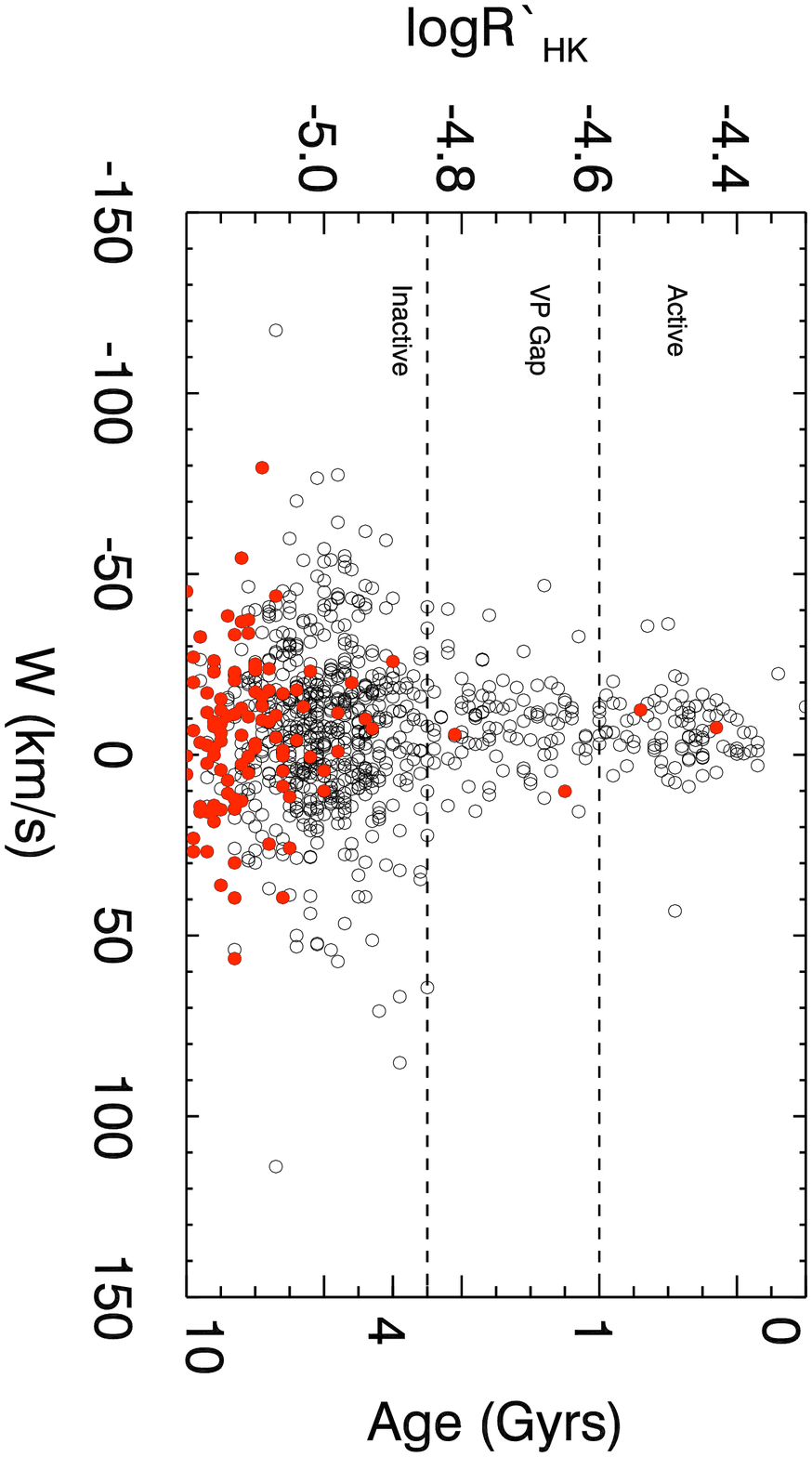}
\includegraphics{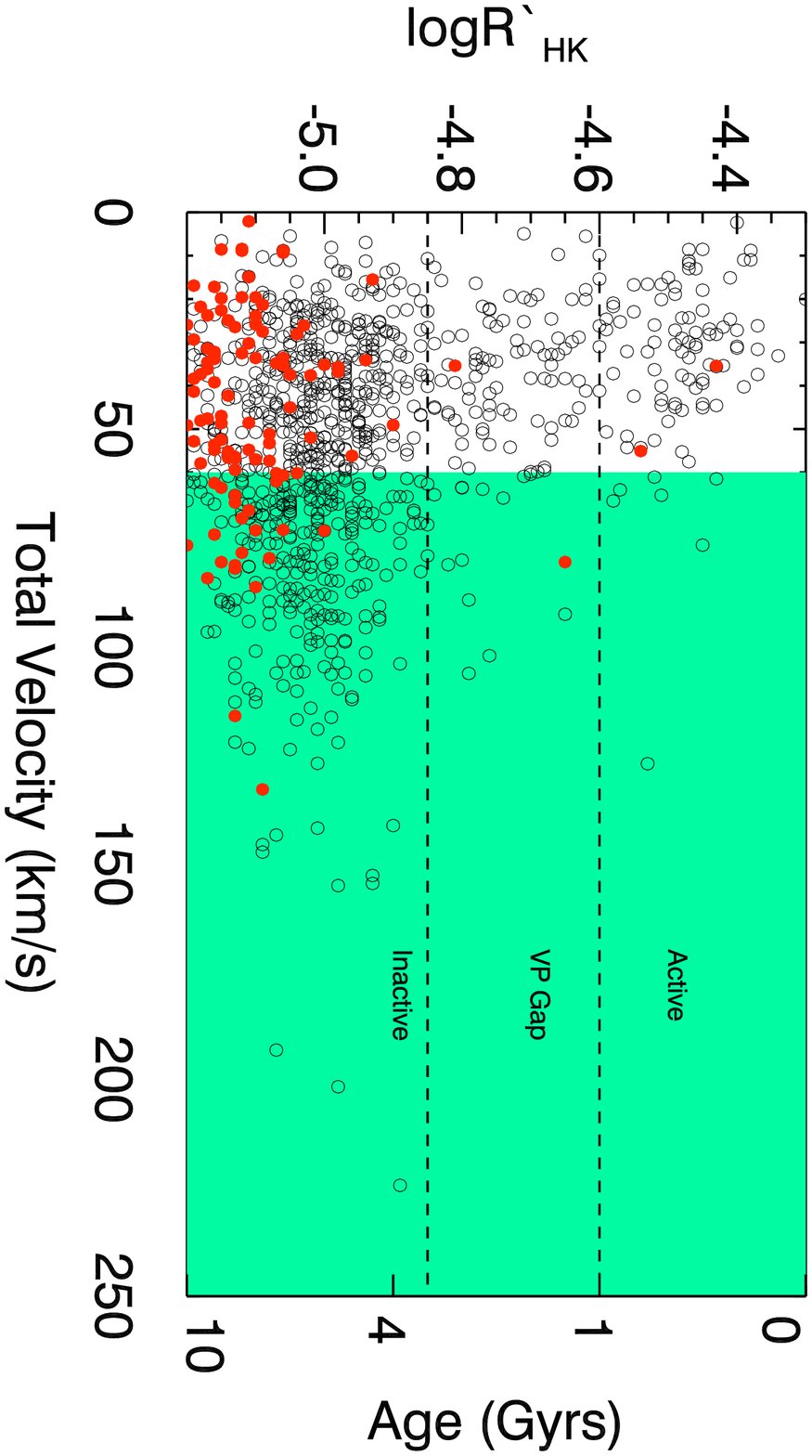}
\vspace{6.0cm}
\caption{The distributions of chromospheric activity against kinematic space motion with main sequence stars as open circles and red filled circles are subgiants.  The bimodal activity 
clusters of stars can be discerned in these plots with a 
cluster of active and inactive stars, bounded by the dashed lines.  Intriguingly the inactive cluster, after the Vaughan-Preston gap, seems to correlate with the dispersing of stars in the 
galactic disk.  Such a result indicates there may be connection between stellar activity, stellar rotation and stellar kinematics.  The green shaded region in the V and total velocity plots 
visually highlights useful parameter space for using kinematics to select inactive stars.}
\label{rhk_kin}
\end{figure*}

An interesting sample distribution to investigate is to study any correlations between the chromospheric activities for solar-type dwarfs 
and SG stars.  Fig.~\ref{rhk_kin} shows all the four kinematic velocities highlighted above plotted against both the sample log$R'_{\rm{HK}}$ and the estimated mean age, which we recall 
was taken from the age-activity relationship of \citet{mamajek08}.  We note that the SG spread in kinematic 
space is no different from MS stars beyond the VP gap region.  We also see that in all panels for each velocity, there is a kinematically condensed region with high activity, 
spreading out into a kinematically broad region at low activity.  In fact, if we plot the upper and lower boundaries of the VP gap region we see this correlates well with these two kinematic 
regions, particularly the step towards a kinematically broad sample for inactive stars.  This step to a kinematically broad sample is apparent in all plots, but is particularly prevalent in the 
V and total kinematic velocity plots.

This result would indicate that there is a statistical correlation between stellar activity and stellar kinematic velocity, whereby the young active stars still retain their birth bulk kinematic 
velocities up to the VP gap, evidenced by the age of the entrance onto the VP gap being around 1~Gyrs which is the age by which stellar kinematics have thermalised enough to 
show no signature from their initial moving group (MG) (e.g. \citealp{clarke10}).  The distributions then show a slow kinematic transition until an age of around 3-4~Gyrs where the 
velocities spread out to fill the kinematic parameter space.  This sudden increase in the spread of velocities also correlates with the end of the VP gap region, telling us there may 
be a correlation between stellar bulk motion and stellar magnetic activity, depending on where the exact edge of the VP gap is located. 

\begin{figure}
\vspace{5.0cm}
\hspace{-4.0cm}
\includegraphics{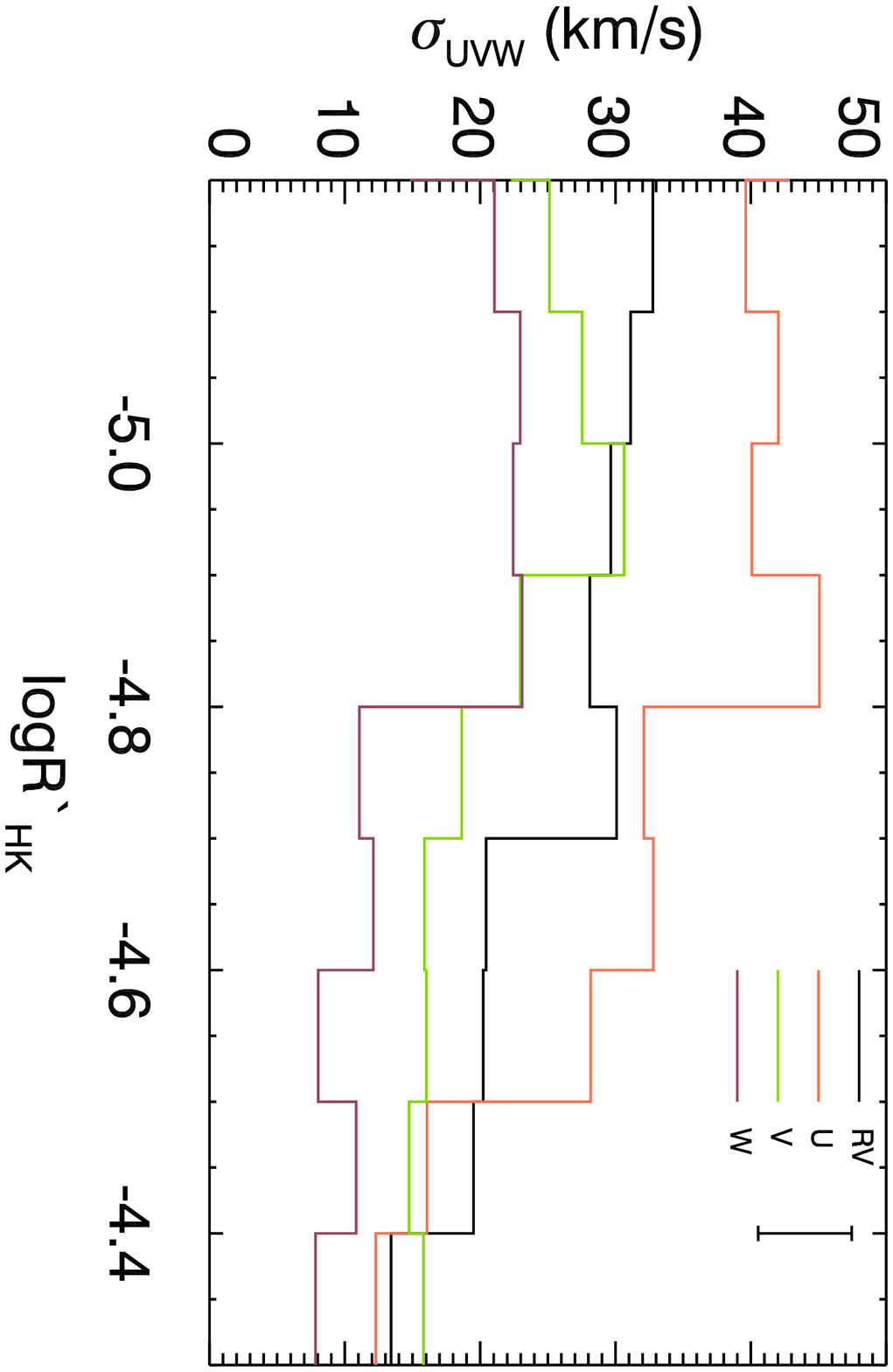}
\includegraphics{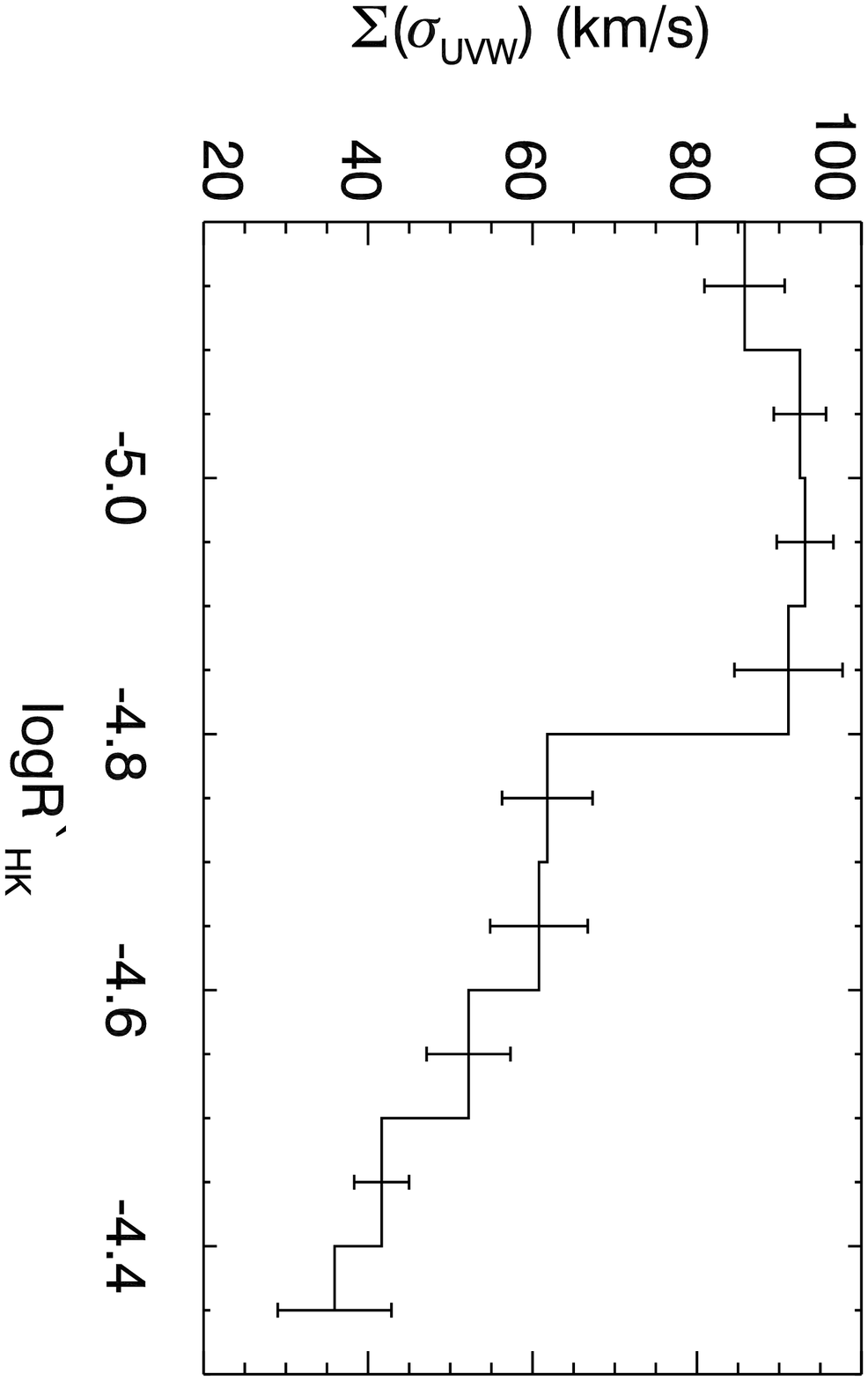}
\vspace{5.0cm}
\caption{The top panel shows histograms of the standard deviation of RV, U, V and W in binned activity regions, along with the average uncertainty.  This plot better highlights the increase 
in the dispersion of kinematic motion vectors as a function of decreasing chromospheric activity, and hence increase in time.  The step up to large dispersion and a flat distribution is seen 
in all histograms.  The lower panel shows the binned totals of the standard deviations for U,V and W space velocities.  This allows a clearer picture of the step after the Vaughan-Preston 
gap region.}
\label{std_kin}
\end{figure}

To better illustrate this result we show Fig.~\ref{std_kin}, where the upper panel shows the standard deviation of the RV measurements along with the U, V and W kinematic 
velocity components and binned as a function of chromospheric activity.  In general for these quantities there is a flat or slow rise in 
the distributions until the lower edge of the VP gap, when there is a step function up to higher dispersion followed by a flat trend.  The lower panel combines all the kinematic standard 
deviation distributions to better highlight this trend and with higher precision which gives three times the spread information per activity bin.  The uncertainties come from Poisson 
statistics and show that the step is significant here at the $\sim$2$\sigma$ level.  Given the apparent strength of this result we seek a model to describe why the chromospheric activity 
is correlated with the kinematic space motion for the older field population of solar type stars i.e. the large increase in space motion for stars with activities below -4.85, and we speculate 
on such a model here.

\subsection{Model Scenarios}

\subsubsection{Magnetic Saturation}
Many previous works have studied the phenomena of saturation of the activity indicators of dwarf stars and models invoking a saturation of the stellar dynamo (\citealp{charbonneau92}), 
a change in the relationship between the activity indicators and the magnetic field strength (\citealp{vilhu84}) or simply that the saturation arises through 
other physical means (e.g. \citealp{solanki97}; \citealp{ryan05}).  Whatever the mechanism that produces the saturation, it is well established that saturation of 
magnetic activity is a key element of stellar evolution.  We can clearly see the presence of saturation in the bottom panel of Fig.~\ref{act_rot} where there 
are many stars with strong magnetic activity that exhibit a range of $v$sin$i$'s.

This saturation, if ubiquitous, could lead to the kinematical distributions we see in Fig.~\ref{rhk_kin} since we could have a pile-up of stars above the VP gap which 
are in the process of spinning down but are still in the regime dominated by the saturation of magnetic activity.  After the stars reach a rotational velocity that is 
below the saturation limit, they then begin the transition across the VP gap region until they reach the inactive phase of their evolution.  Given that it takes 
a long time to drive a star's rotational velocity from the typical rotation velocity at a log$R'_{\rm{HK}}$ of -4.85 ($\sim$3kms$^{\rm{-1}}$) down to essentially zero, there 
is a pile-up of stars with weak magnetic fields and hence inactive chromospheres, somewhat conforming to the distributions we see Fig.~\ref{rhk_kin} if the saturation 
limit occurs across a spread of different ages because the stars begin their life with a large spread of rotational velocities.  \rm

\subsubsection{Residual Molecular Gas}
From the age estimates in Fig.~\ref{rhk_kin} we see that the end of the VP gap and the kinematic scattering first occur at ages around 3-4~Gyrs.  
If we look at the beginning of the VP gap, we see this occurs at ages around 1-2~Gyr or so.  Since this is around the age of the oldest MGs (e.g. HR1614 \citealp{desilva07}), we might envisage 
that this active subset are artificially 
held in a fast rotating active state through accretion of small amounts of residual gas from the remnant molecular cloud from which they formed.  The stars that have moved far enough 
away from their point of origin are not embedded in any gas and they quickly transition across the VP gap.  Such a scenario would not require any complex changing of the magnetic 
dynamo since the active bump is simply an artifact of an external source driving the rotational dynamo within.  This would suggest that the active part of the bimodal activity distribution represents 
the actual spread of zero-age main sequence rotational velocities and after the gas has fully dissipated the rotational energy is no longer available from accretion and the velocities begin to decay.

The inactive bump then represents the pile-up of objects after they have quickly transitioned from a magnetically active state to a magnetically quiescent state.  The timescale 
between when a star is free from any residual gas from its stellar nursery and when it is kinematically thermalised due to disk heating mechanisms is similar (\citealp{desimone04}).  
This may indicate that disk heating 
has a direct effect on stellar magnetic fields.  It could be that once a star is free from the residual molecular cloud gas from which it formed, the time for transition to the inactive phase of it's 
activity lifetime is a direct consequence of the stellar disk environment in which it moves through.  The spiral density waves and gravitational potentials within the galaxy 
(e.g. \citealp{desimone04}) actually work to slow a star's rotational velocity and hence drive a less powerful magnetic field.

\subsubsection{Planetesimal Accretion}

We discuss this low level accretion of gas in terms of the residual molecular gas since results from the Spitzer Space Telescope suggest that beyond an age of $\sim$10~Myrs there are no 
primordial disks remaining around weak-lines T-Tauri stars (\citealp{wahhaj10}).  However, there is a transition between primordial disks and the onset of debris disks which are 
found around 10-20\% of main sequence stars from Spitzer studies (e.g. \citealp{silverstone06}; \citealp{trilling08}).  This could indicate that stars are being held at high rotation early in their life 
due to the accretion of planets or rocky debris material leftover from the planet formation process.  However, to hold stars at high rotation for so long, probably requires a dynamical scenario 
different from the proposed planet migration process through disk migration (\citealp{lin}), given the disks are gone.  Possibly a planet-planet scattering like scenario could yield the 
necessary accretion required (\citealp{rasio96}).

The planet-planet scattering process in a gas free environment can explain some interesting properties of the current distribution of exoplanets (\citealp{chatterjee08}) and observationally 
it does appear to play some part in sculpting planetary systems (\citealp{triaud10}).  If this is the case, and since we know planet formation is an integral part of the star formation and 
early stellar evolution process, then it stands to reason that dynamical formation of planets will affect their host stars to some degree.  An example of this is the current model proposed 
by \citet{bouvier08} to explain the observed depletion of lithium in planet host stars (\citealp{gonzalez00}; \citealp{israelian09}).  Therefore, it could be that stars with ages below 
$\sim$1~Gyr are accreting planets/planetesimals, transferring some angular momentum onto the host star and once the system is dynamically stable the star is free to rotationally stabilise 
and follow the evolution of the other two scenarios mentioned above.  If this scenario is true 
then it allows us to put observational limits on the timescale by which planet-planet scattering models must work and this timescale agrees well with the timescale of $\sim$10$^{8}$~yrs for 
planet-planet scattering set out in \citet{matsumura08}.  

We also briefly mention to what degree we might expect planetesimal accretion from the current ensemble of detected planets.  \citet{howard10} recently analysed the Keck sample 
of 166 stars and have suggested that 1.2$\pm$0.2\% of stars host hot-Jupiters with orbital periods less than 50~days.  However, once we look at planets with masses between 0.5-2~M$_{\oplus}$ 
then the occurrence rate is found to be 23$^{\rm{+16}}_{\rm{-10}}$\% orbiting Sun-like stars in this period domain, all following a power law described by d$f$/dlogM = 0.39M$^{\rm{-0.48}}$.  Previous 
studies of chromospheric activity have shown the fraction of active stars in a solar-type sample in the solar neighbourhood is found to be $\sim$27\% (see Henry et al. Fig.~7).  
This number is in stark agreement with the Keck number of low mass bodies on short period orbits and could be pointing towards planetesimal accretion as being the cause of the active population 
of solar-type stars in the solar neighbourhood if we assume a large number of these bodies were brought to their current orbits by dynamical processes.  This would only make sense if 
the other $\sim$70 percent without short period planets to disrupt or accrete were already further down the activity evolution scale, meaning the age-activity correlation would be affected by 
introducing a large scatter into this relationship.  Although it could be that these other stars did have planets/planetesimals yet they are gone since they were accreted, ejected or disrupted by 
dynamical interactions.

\citet{guillochon10} have modelled multi-orbit encounters of planetary bodies with masses M$_{\rm{p}}$$\ge$0.1~M$_{\rm{J}}$ including tidal effects and have shown that ejection, 
disruption or collision of planetary bodies can inject a large fraction of angular momentum onto the host star, depending on the initial orbital radii and eccentricity.  Their simulations show 
that ejected planets can impart almost half of their mass onto the central star, completely disrupted planets can also impart around half of their mass onto the star and accreted planets 
can impart anywhere from half to all of their mass onto the host star.  They find that ejection or disruption can deposit a Sun's worth of angular momentum onto the host star and some 10\% of 
stars in their simulations are found to have three times the solar spin rate after disruption.  Given complete accretion is found to occur twice as often as ejection or disruption then clearly the infall 
of planetary/rocky material can have a significant affect on the stellar spin rate and the timescale whereby such dynamical evolution works is in agreement with the timescales where major changes 
occur in the chromospheric activity distribution i.e. the beginning of the VP gap region.

In addition to the model scenarios we lay out above, we highlight more practically just how kinematics can be used to select good planet search candidate stars.  The green regions in the 
V (top right) and total velocity planets (bottom right) are regions bounded by velocities of $\le$-50kms$^{\rm{-1}}$ and $\ge$70kms$^{\rm{-1}}$ respectively.  Simply selecting MS or SG 
stars with such velocities allows a high fraction of inactive stars to be selected, 
stars which give the best for precision RV surveys.  The fraction of stars in the kinematic V plane less than or equal to -50kms$^{\rm{-1}}$ and below the VP gap is over 93\%, and 
for the total velocity we find a similar value for stars greater than or equal to 70kms$^{\rm{-1}}$.  This relates to 23\% and 28\% of the total number of stars beyond the VP gap region, a high fraction 
considering the number of such stars in the Hipparcos or Tycho catalogues.  If we extend the total velocity value down to 60kms$^{\rm{-1}}$ we then pick up 45\% of the total number of inactive stars 
but with 12\% contamination of stars above the VP gap.  The value of this selection is that the V component does not require any spectra and can be ready measured from the Hipparcos 
catalogue and also the total velocity component can be useful for planet search surveys, like the AAPS (\citealp{tinney02}; \citealp{jenkins06c}), that do not cover the CaHK region as part 
of their doppler spectra.

\subsection{Disk Heating}

\begin{figure}
\vspace{5.0cm}
\hspace{-4.0cm}
\includegraphics{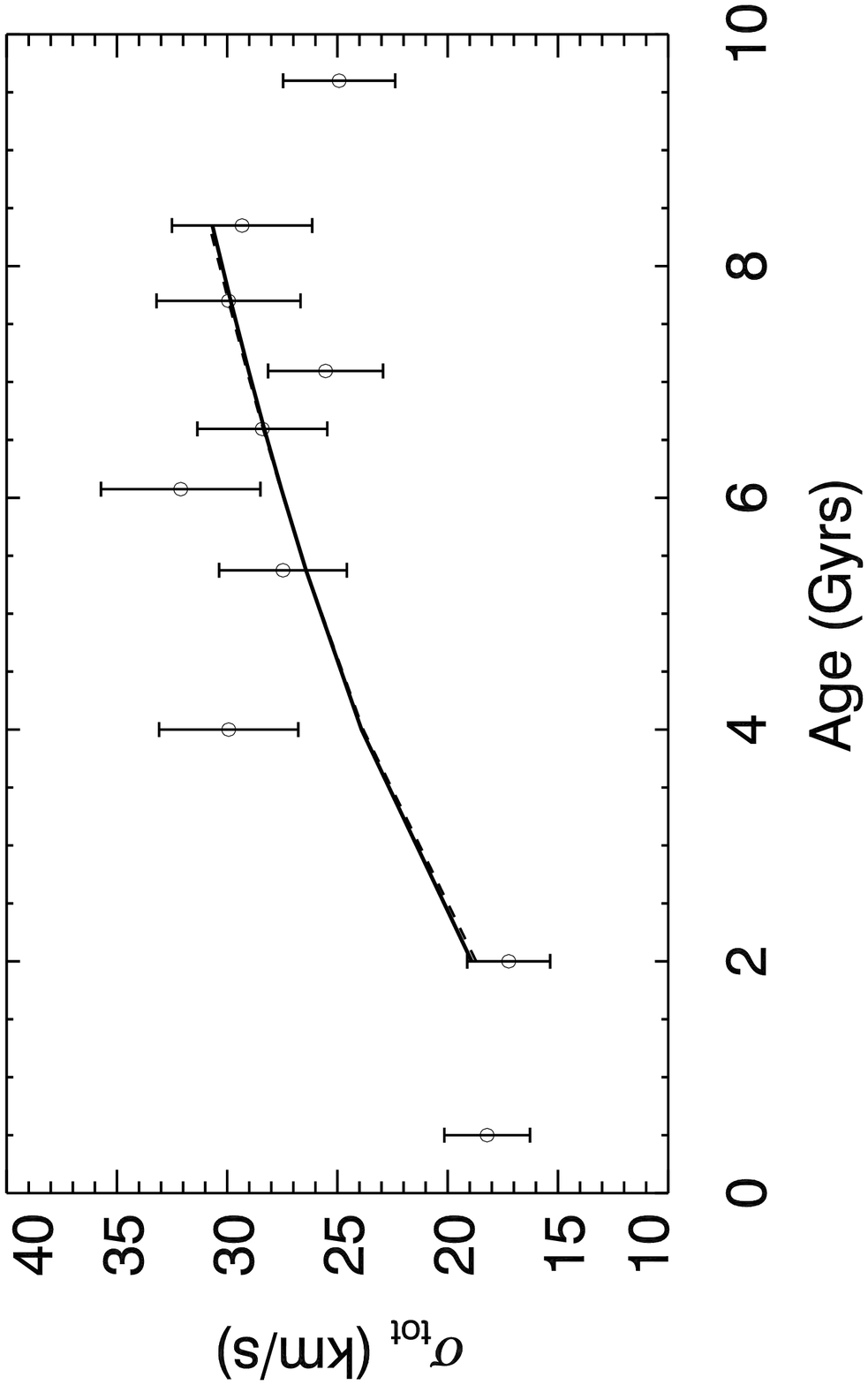}
\includegraphics{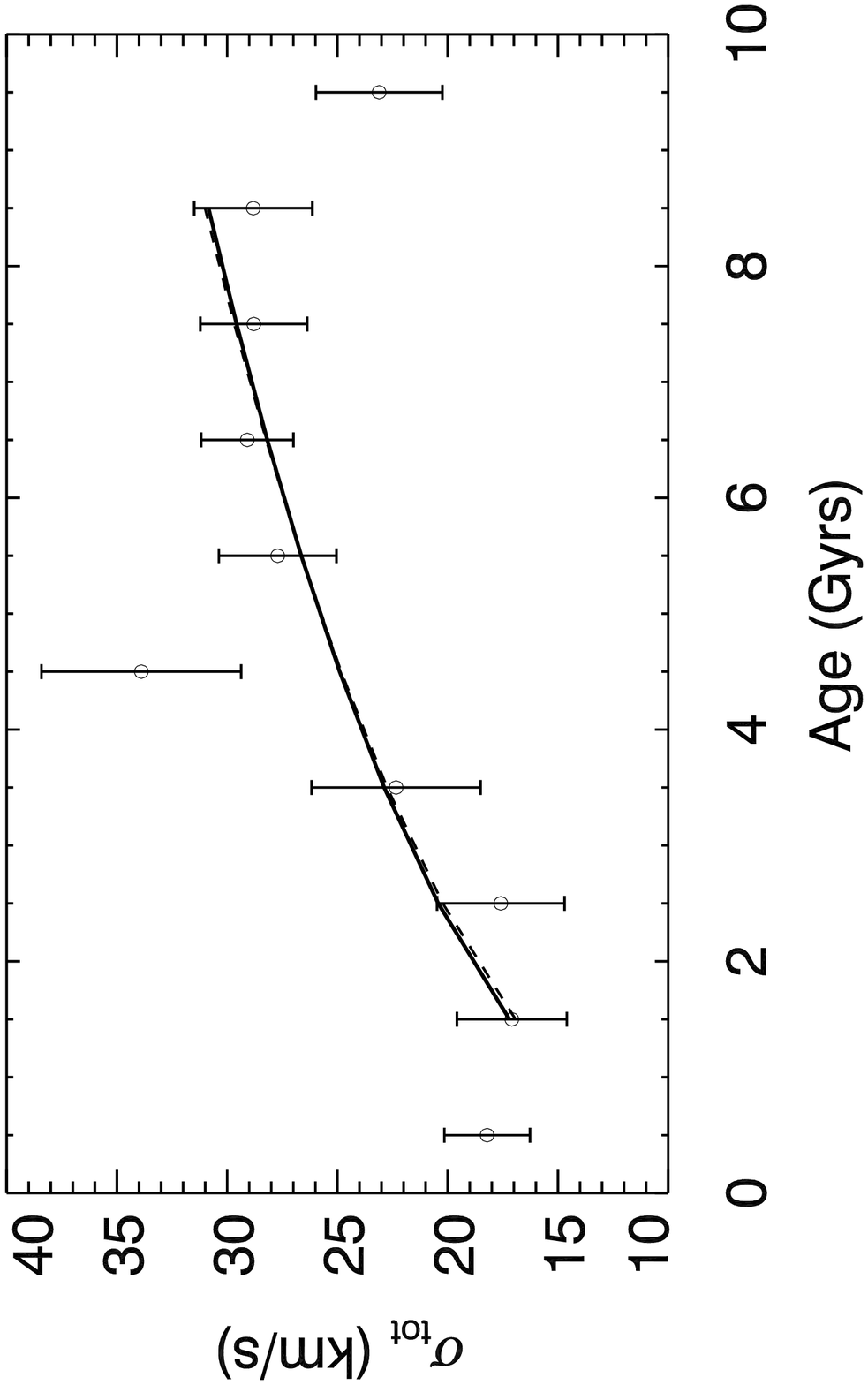}
\vspace{5.0cm}
\caption{Binned total kinematic velocities for our sample as a function of age derived using the chromospheric activities.  The uncertainties are from Poisson statistics.  
The top panel shows the fit where the bins contain equal numbers of stars and the lower panel is binned in 1~Gyr intervals.  The solid curve represents the best fit power law, 
parameterising the disk heating exponent as a function of time.  The dashed line is the best fit theoretical disk heating model with a power law index of 0.349 (\citealp{aumer09}).}
\label{heating}
\end{figure}

In Nordstrom et al. and Holmberg et al. a magnitude limited effort was made to better constrain the disk heating power law function, a fundamental parameter that governs the kinematic 
populations of stars in the galaxy.  As large structures like giant molecular clouds and spiral arms heat the disk of the galaxy by adding gravitational potentials to the disk, this energy 
is transferred to kinetic energy in the form 
of stellar space velocities and hence studying the age-velocity relationship (AVR) can lead to a constrained empirical measurement for this parameter.  Nordstrom et al. found the parameter to 
be 0.34$\pm$0.05, whereas Holmberg et al. found a value of 0.40, both in close agreement with the predicted theoretical value of 0.33$\pm$0.03 from \citet{binney00} and the latest 
prediction of 0.349 (\citealp{aumer09} using the new Hipparcos reduction (\citealp{leeuwen09}) and updated isochrones.  

All above mentioned disk heating exponents were derived from isochrone fitting, however in the past, AVR's drawn from chromospheric activity values did not give heating functions 
in good agreement with those predicted from models.  \citet{hanninen02} show a list of values for the heating function and the two drawn from activity-age relations were found to be fairly 
divergent from each other and not in agreement with the isochrone fitting procedure.  Values of 0.26$\pm$0.01 and 0.59$\pm$0.04 were found using chromospheric activities from 
Hanninen et al.'s own group and \citet{fuchs01} respectively.  However, the 0.26 value was dependent on binning of the histogram, which they set to ensure the bins contained around the 
same number of stars.  In Fig.~\ref{heating} we show our fit to the AVR where we plot the total velocity dispersion against time using two separate binning methods.

We utilised all our dataset in this analysis and, as before, made use of the latest age-activity relationships to generate our empirical AVR.  The top panel shows the data for an even 
number of stars in each bin and the lower panel is the data binned in time with bin widths of 1~Gyr.  The solid best fit 
power laws give rise to a disk heating function of 0.331$\pm$0.048 and 0.337$\pm$0.045 for the top and bottom panels respectively, in good agreement with the values derived using 
isochrone fitting.   We also plot on both the latest best fit theoretical model from Aumer \& Binney (dashed curves), where we set the power law index to their derived value of 
0.349 and then minimise the offset.  We clearly see good agreement between our empirical value and that found by theory.

We now note some caveats.  The fits we have shown in Fig.~\ref{heating} relate to only ages between 1-9~Gyrs, where we have dropped out the outer most bins.  We introduced this drop out 
since we have a number of biases and incompleteness in this analysis.  We assume a reddening function equal to zero for all of our sample which will move the position of stars 
on the HR-diagram a little in our selection given a large fraction of our sample are located at distances greater than 70pc.  For instance, this will shift the position of Parenago's 
discontinuity (\citealp{parenago50}); the discontinuity being the sharp end of the increase in total velocity dispersion with $B-V$ colour and hence age.  The youngest stars are those just 
evolved onto the MS and hence reddening should have an adverse affect on these with the introduction of stars not yet evolved onto the MS.  Conversely, the same for the oldest and reddest 
stars in our sample, which will be affected most by the lack of reddening correction.  We plan to revisit the AVR issue in the future since we are currently in the process of refining a new 
method for measuring precise atomic abundances for these FEROS stars by spectral synthesis fitting (see \citealp{pavlenko10}) 
and since the distribution functions of stellar populations are related to properties such as metallicity and age (\citealp{binney10}) we can re-evaluate the AVR as a function 
of atomic abundance and try to constrain the dependence of the disk heating parameter on such properties.

\section{Summary}

We have spectroscopically analysed a sample of southern FGK MS dwarfs and SG stars to better understand stellar chromospheric activities and kinematic space motions.  In 
the process of this we studied a number of ways to calibrate activities (log$R'_{\rm{HK}}$) onto the commonly used Mt.~Wilson system of measurements and found that for resolving 
powers above 
$\sim$10$'$000-20$'$000 the rms scatter around a linear calibration is essentially the same.  Also it only really escalates for resolving powers below $\sim$2$'$500, meaning even moderate 
resolutions can 
provide highly accurate log$R'_{\rm{HK}}$ indices.  In addition, we examined the need to fully mimic the Mt.~Wilson setup synthetically on 1D spectroscopic datasets and found that 
changing from triangular Ca \sc II  \rm HK core bandpasses to square core bandpasses is negligible at moderate to high resolution.

Next we compared our new activities to those already in the literature to look for any zero point offsets.  Given we have attempted to calibrate our activities using long term stable 
stars as our standards, such a test should provide a robust examination of drift changes inherent in older activity studies.  We find no offset between our dataset and that of \citet{wright04}, 
however we see a small indication of an offset between our data and that from \citet{gray05} and a significant offset between our data and that from \citet{henry96}.  It seems clear that the 
offset between our data and that of Henry et al. is due to drift of the calibration stars used in Henry et al. as we found no offset against Henry et al. in Paper I when we calibrated using similar 
calibrators to Henry et al.  We note that this point should be revisited in the future with a larger sample of long term stable calibrator stars.

We also looked at both MS stars and SG stars separately and could show that the age-activity relationship does indeed continue for these stars which are older than the Sun.  
When we look at the two populations we can show they are not drawn from the same parent population and also the mean differences between the samples indicate that these SGs are 
around 2.5~Gyrs older on average.  Also the SG distribution and the MS star distributions have different Gaussian widths.  This indicates that when stars transition to the SG 
branch and their convective envelopes deepen then their activities stabilise and hence their cycles are minimised.  An alternate possibility is that there is a pile-up around a minimum log$R'_{\rm{HK}}$ 
which does not change much since the rotational velocity is much lower and approaching zero.  Also there is a small indication of an increased spin-down timescale for K dwarfs when 
compared to earlier F and G stars.

We looked at the kinematic space motions of our sample and in particular how the kinematic distributions look as a function of chromospheric activity.  There appears to be 
a correlation between activity and kinematic space motion, however it is difficult to rule out a third party correlation factor here.  If these two are indeed correlated we speculate on models to 
explain this, from saturation of the activity indicators to influences by external sources such as accretion of residual gas from the formation of the star cluster, galactic potentials such 
as spiral density waves or accretion of planets or planetesimals.

Finally, we applied the latest age-activity relations for our sample to empirically constrain the disk heating parameter from the age-velocity relationship.  We find the best fit power law to our 
large data set as having an exponent of 0.337$\pm$0.045, in excellent agreement with values derived from isochrone fitting and theoretical models.


\begin{acknowledgements}

JSJ acknowledges funding by Fondecyt through grant 3110004 and, along with FM, PR, ADJ and MTR, partial support from Centro de Astrof\'\i sica FONDAP 15010003, the  GEMINI-CONICYT FUND 
and from the Comit\'e Mixto ESO-GOBIERNO DE CHILE.  PR also acknowledges support from Fondecyt 11080271.  FM and DJP are supported by RoPACS, a Marie Curie Initial Training Network 
funded by the European Commission's Seventh Framework Program.  ADJ achnowledges support by Fondecyt grant 3100098.  JRAC is supported by GEMINI-CONICYT FUND No. 32090002 and 
also by the Mileno - MIDEPLAN ICM Nucleus P07-021-F.

\end{acknowledgements}

\bibliographystyle{aa}
\bibliography{refs}

\begin{sidewaystable*}
\large 
\centering 
\label{tab:act_cals}
\vspace{15.0cm}
\caption{List of FEROS activity calibrators}
\begin{tabular}{ccccccccccc}
\hline 
\multicolumn{1}{c}{Star}  & \multicolumn{1}{c}{$B-V$} & \multicolumn{1}{c}{V} & \multicolumn{1}{c}{S$_{\rm{FEROS}}$}  &  \multicolumn{1}{c}{S$_{\rm{MW}}$}  & \multicolumn{1}{c}{log$R'_{\rm{HK}}$} & 
\multicolumn{1}{c}{$v$sin$i$ (kms$^{\rm{-1}}$)} & \multicolumn{1}{c}{RV (kms$^{\rm{-1}}$)}  & \multicolumn{1}{c}{U (kms$^{\rm{-1}}$)} & \multicolumn{1}{c}{V (kms$^{\rm{-1}}$)} &  \multicolumn{1}{c}{W (kms$^{\rm{-1}}$)} \\ \hline
& & & & & & & & & & \\
Baliunas et al. (1995)  & & & & & & & & & & \\
----------------  & & & & & & & & & & \\
HIP13702    & 0.471   & 5.58  & 0.137  & 0.172$\pm$0.004  & -4.87     &  --   &    --   &    --   &    --   &    -- \\
HIP69536    & 0.537   & 5.53  & 0.105  & 0.136$\pm$0.003  & -5.16     & 4.9   & -39.8$\pm$1.7   & -39.0$\pm$0.9   & -32.5$\pm$0.8   & -26.0$\pm$1.6 \\
HIP112935  & 0.487   & 5.16  & 0.113  & 0.145$\pm$0.004  & -5.06     & 3.9   &  11.4$\pm$1.7   & -58.1$\pm$1.2   &  -6.9$\pm$1.3   & -33.8$\pm$1.3 \\
HIP112935  & 0.480   & 5.16  & 0.111  & 0.142$\pm$0.004  & -5.09     & 3.9   &  11.4$\pm$1.7   & -58.1$\pm$1.2   &  -6.9$\pm$1.3   & -33.8$\pm$1.3 \\
\hline 
\end{tabular}
\medskip

Note the overlapping calibrators between the Baliunas et al. and Lockwood et al. studies and the multiple observations. 

\end{sidewaystable*}

\longtab{4}{
\begin{longtable}{cccccccccc}
\caption{\label{tab:act_kinematics} FEROS activities, rotational velocities, radial-velocities and kinematic space motions.}\\
\hline\hline
Star  &  $B-V$ &  V &  S$_{\rm{MW}}$  &  log$R'_{\rm{HK}}$ &  $v$sin$i$ &  RV  &  U &  V &   W \\
   &   &  mags &    &   &  kms$^{\rm{-1}}$ &  kms$^{\rm{-1}}$  & kms$^{\rm{-1}}$ &  kms$^{\rm{-1}}$ &  kms$^{\rm{-1}}$ \\
\hline
\endfirsthead
\hline
 Star  &  $B-V$ &  V &  S$_{\rm{MW}}$  &  log$R'_{\rm{HK}}$ &  $v$sin$i$ & RV  &  U &  V &   W \\
\hline
\endhead
\hline
\endfoot

HIP72 &   0.671 &    8.97 &   0.146$\pm$0.004 &   -5.10 &      2.1 &    -52.1$\pm$1.6 &    -48.9$\pm$5.4 &    -97.9$\pm$9.6 &     19.8$\pm$3.8 \\
HIP81 &   0.642 &    8.57 &   0.168$\pm$0.004 &   -4.94 &      2.0 &     -0.2$\pm$1.6 &     49.6$\pm$2.7 &    -10.8$\pm$0.9 &     -5.7$\pm$1.5 \\
HIP135 &   0.620 &    8.64 &   0.195$\pm$0.004 &   -4.80 &      2.2 &     -7.8$\pm$1.6 &      9.2$\pm$0.6 &    -26.1$\pm$1.6 &     -6.3$\pm$1.6 \\
HIP599 &   0.599 &    8.23 &   0.157$\pm$0.004 &   -5.00 &      1.2 &     10.1$\pm$1.6 &     15.8$\pm$0.9 &     11.1$\pm$0.8 &     -7.9$\pm$1.6 \\
HIP772 &   0.687 &    8.52 &   0.152$\pm$0.004 &   -5.06 &      2.1 &    -26.1$\pm$1.6 &      3.9$\pm$0.6 &     -6.7$\pm$0.4 &     27.6$\pm$1.6 \\
HIP846 &   0.578 &    8.92 &   0.165$\pm$0.004 &   -4.94 &      1.0 &    -12.9$\pm$1.6 &    -67.8$\pm$6.4 &    -12.1$\pm$0.9 &      8.1$\pm$1.6 \\
HIP1000 &   0.707 &    8.73 &   0.175$\pm$0.004 &   -4.94 &      7.0 &     34.5$\pm$1.8 &    -40.5$\pm$3.8 &    -51.9$\pm$5.0 &    -46.8$\pm$2.1 \\
HIP1173 &   0.657 &    9.18 &   0.172$\pm$0.004 &   -4.93 &      1.7 &     18.8$\pm$1.6 &     -1.8$\pm$0.7 &    -24.1$\pm$2.2 &    -16.8$\pm$1.6 \\
\end{longtable}

}

\end{document}